\documentclass[prd,twocolumn,superscriptaddress,showpacs,nofootinbib, amsmath, amssymb]{revtex4-1}

\usepackage{graphicx}
\usepackage{hyperref}
\usepackage{color}
\usepackage[utf8]{inputenc}
\usepackage{scalefnt}
\usepackage{lineno}
\usepackage{here}

\begin{document}

\title{Dependence of accessible dark matter annihilation cross-sections on the density profiles of dwarf spheroidal galaxies with the Cherenkov Telescope Array}

\author{Nagisa Hiroshima}
\affiliation{RIKEN Interdisciplinary Theoretical and Mathematical Sciences (iTHEMS),
Wako, Saitama 351-0198, Japan}
\affiliation{Institute for Cosmic Ray Research, The University of Tokyo,
Kashiwa, Chiba 277-8582, Japan}
\affiliation{Institute of Particle and Nuclear Studies, KEK, Tsukuba,
Ibaraki 305-0801, Japan}
\author{Masaaki Hayashida}
\affiliation{Department of Physics, Faculty of Science and Engineering, Konan University, 8-9-1 Okamoto, Kobe, Hyogo 658-8501, Japan}
\author{Kazunori Kohri}
\affiliation{Institute of Particle and Nuclear Studies, KEK, Tsukuba,
Ibaraki 305-0801, Japan}
\affiliation{The Graduate University for Advanced Studies (SOKENDAI), Tsukuba, Ibaraki 305-0801, Japan}
\affiliation{Rudolf Peierls Centre for Theoretical Physics, The University of Oxford, Parks Road, Oxford, OX1 3PU, UK}

\begin{abstract}
Dwarf spheroidal galaxies are excellent targets in $\gamma$-ray searches for dark matter. We consider dark matter searches in dwarf spheroidal galaxies (dSphs) with the Cherenkov Telescope Array (CTA). The aim of this work is to reveal a quantitative and precise dependence of the accessible dark matter annihilation cross-sections on the dark matter density profiles of dSphs and on the distance to them. In most data analyses, researchers have assumed point-like signals from dSphs because it is difficult to resolve the expected emission profiles with current $\gamma$-ray observatories. In future however, CTA will be able to resolve the peak emission profiles in dSphs. 
 We take several variations of the dark matter density profile of Draco dSph as examples and analyze the simulated  observations of with CTA. We derive the accessible region of the dark matter annihilation cross-section with each dark matter density profile. The accessible region of the annihilation cross-section can differ by a factor of 10 among plausible profiles. We also examine the dependence on the distance to the target dSphs by assuming the same profiles of dSphs at different distances. Closer targets are better due to the higher J-factor, while their spatial extension significantly degrades our reach to the annihilation cross-section compared to the value expected from a simple distance-scaling of the J-factor. Spatial extension of the source affects the probable parameter region in energy-dependent ways. In some $\gamma$-ray energy ranges, this behaviour becomes moderately dependent on the properties of the observation facility.

\end{abstract}

\date{\today}

\maketitle

\section{Introduction}
\label{sec:intro}

Dark matter (DM) is a massive and invisible matter-component of the
Universe
\cite{Bergstrom:2000pn,Bertone:2004pz,Gaskins:2016cha}. Rotation
curves of galaxies \cite{Zwicky:1933gu,Zwicky2009,vanAlbada:1984js,
  Salucci:2002jg} and bullet-cluster like encounters
\cite{Barrena:2002dp, Clowe:2003tk} are examples that indicate the
existence of DM. Standard cosmology also requires DM, since
non-relativistic matter components different from baryons are
necessary to form structures of the Universe
\cite{Peebles:1982ff,Ade:2015xua}.  Cosmological observations indicate
that DM occupies approximately a quarter of the total energy in the
Universe~\cite{Komatsu:2010fb,Ade:2015xua,Akrami:2018vks,Aghanim:2018eyx}.

Varieties of candidates for DM are proposed. 
One possibility is that  DM is a new particle:  weakly interacting massive particles (WIMPs) (e.g. ~\cite{Bringmann:2006mu,Rott:2012gh}), strongly interacting massive particles (e.g.~\cite{Mohapatra:1999gg,Hochberg:2014dra}), axions (e.g.~\cite{Preskill:1982cy,Rosenberg:2000wb,Visinelli:2009zm}), or sterile neutrinos (e.g.~\cite{Dodelson:1993je,Shi:1998km,Abazajian:2001nj,Boyarsky:2009ix}) are examples. Non-particle solutions like primordial black holes (e.g.~\cite{Afshordi:2003zb,Carr:2016drx,Carr:2016hva,Carr:2018rid,Kohri:2018qtx}) are also considered. In this paper, we focus on DM categorized as WIMPs. 
WIMPs are one of the best-studied candidates proposed in theories beyond the standard model like supersymmetric extensions (e.g.~\cite{Haber:1984rc,Jungman:1995df,Edsjo:1997bg,Feng:2000gh,Giudice:2004tc}). For WIMPs to be DM particles that explain non-relativistic, electromagnetically neutral, invisible components in \cite{Ade:2015xua,Akrami:2018vks}, their mass must be around  $m_{\rm DM}\sim{\cal O}$(GeV) to ${\cal O}$(TeV), and have a velocity-averaged freeze-out annihilation cross-section $\left<\sigma v\right>\sim3\times10^{-26}$ cm$^3$/s \cite{Steigman:2012nb}. This value of cross-section is referred to as the canonical cross-section.

WIMPs as DM can be detectable through their feeble interaction with standard model particles. Three kinds of strategies are pursued: productions of DM with colliders (e.g.~\cite{Aaboud:2017aeu,Sirunyan:2017xgm}); measuring the scattering between DM particles and nuclei (e.g.~\cite{Akerib:2016vxi,Amole:2017dex,Aprile:2018dbl}), called direct detection experiments; and the search for standard model particles produced after DM self-annihilation in the Universe, called indirect detection experiments. There has been no confirmed detection of particle DM neither DM yet. For WIMP models of $m_\mathrm{DM}\sim{\cal O}(1-10)$ GeV, $\gamma$-ray observations already constrain the DM annihilation cross-section to be smaller than the canonical value ~\cite{Fermi-LAT:2016uux}. Lighter DM is constrained from structure formation (e.g ~\cite{Tremaine:1979we,Abazajian:2005xn,Horiuchi:2013noa,Boyarsky:2008xj}). WIMPs heavier than $m_\mathrm{DM}\sim{\cal O}(10)$ GeV are less constrained and expected to be discovered or excluded in ongoing and future experiments.

Indirect detection experiments have advantages in DM searches at higher energy ranges of $m_{\rm DM}\gtrsim{\cal O}(1)$ TeV. Techniques for astrophysical observations to detect high-energy emissions  are already developed~\cite{Verzi:2017hro,Aab:2018chp,Aartsen:2017sml,Abeysekara:2017hyn,Acharya:2017ttl}. A plethora of projects searching DM signals in the Universe with charged cosmic-rays (e.g. ~\cite{Aguilar:2015ooa,TheDAMPE:2017dtc,Motz:2015cua}), neutrinos (e.g. ~\cite{Aartsen:2017ulx,Albert:2016dsy}) and $\gamma$-rays  are ongoing. In general, astrophysical emissions dominate over DM signals, and elaborate strategies are required in the indirect DM search. 
Spectral and morphological information of emissions help to identify the sources. Considering DM searches in $\gamma$-rays with a facility of threshold energy $E_{\rm th}$, the flux from DM annihilations is
\begin{equation}
\label{eq:gammaflux}
\phi=\frac{1}{4\pi}\frac{\left<\sigma v\right>}{2m_\mathrm{DM}^2}\int^{m_\mathrm{DM}}_{E_\mathrm{th}}dE\frac{dN_\gamma}{dE}\cdot \ J
\end{equation}
where
\begin{equation}
\label{eq:Jfactor}
J=\int d\Omega \frac{dJ}{d\Omega}=\int d\Omega \int ds \  \rho_\mathrm{DM}^2.
\end{equation}
In Eq.~(\ref{eq:gammaflux}), all quantities except for $J$ are determined from particle physics. The part shown as $J$ in Eq.~(\ref{eq:Jfactor}) is referred to as the ``(astrophysical) J-factor". Since the J-factor is defined as the line-of-sight integral over the squared DM density $\rho_{\rm DM}^2$, the signal sensibly depends on the density profile and precise information about the DM distribution at the source is necessary to reliably derive the WIMP properties. The distribution of the DM is determined from stellar kinematics in optical observations (e.g.~\cite{Posti2019}).

The Galactic center is considered as one of the best targets to search DM signals in $\gamma$-rays (e.g. ~\cite{Gondolo:1999ef}) because it is expected to have the highest J-factor among known targets with $J\sim{\cal O}(10^{21-22})$ GeV$^2$ cm$^{-5}$. 
Attentive strategies in the separation of DM signals from astrophysical emissions are required since the galactic center is very bright in astrophysical $\gamma$-ray emissions~\cite{TheFermi-LAT:2017vmf}. Also, the determination of the precise shape and a normalization of the DM density distribution at the very center of the Milky Way galaxy are remaining issues~\cite{HESS:2015cda,Pierre:2014tra,Gammaldi:2016uhg,Taylor:2015jaa}. Dwarf spheroidal galaxies (dSphs) are satellites of the Milky Way galaxy and also good regions to focus on as first pointed out by Ref.~\cite{Lake:1990du} and later in Ref.~\cite{Evans:2003sc}. They are spatially extended objects of ${\cal O}(1)$ degrees located in high-latitude regions of the Milky Way galaxy. Several tens of dSphs are already identified with available stellar kinematics data, and the number of confirmed dSphs is continuously increasing~\cite{Martin:2015xla,Kim:2015ila,Laevens:2015una,Laevens:2015kla,Luque:2015txp,Koposov:2015jla,Bechtol:2015cbp,McConnachie:2012vd,Gaia2018,Simons2018}. Stellar motions in dSphs indicate that they are dense and DM dominated objects ~\cite{Strigari:2006rd,Charbonnier:2011ft,Hayashi:2016kcy} with mass-to-luminosity ratios reaching $\sim 10^3M_\odot/L_\odot$~\cite{Mateo:1998wg,Strigari:2007at,Battaglia:2013wqa,Pace:2018tin,Irwin:1995tb}. No significant $\gamma$-ray emissions have been confirmed in dSphs although possibilities that some of them contain $\gamma$-ray sources could not be excluded~\cite{Geringer-Sameth:2015lua,Fermi-LAT:2016uux}.

Stacking analyses on dSphs by the ${\it Fermi}$ collaborations give the tightest upper limits on DM annihilation cross-sections~\cite{Abdo:2010ex,Ackermann:2011wa,Ackermann:2013yva,Ackermann:2015zua, Charles:2016pgz, Fermi-LAT:2016uux}.
For DM of $m_{\rm DM}$ $\lesssim$ ${\cal O}(100)$ GeV, the upper limits already reach to the canonical value ~\cite{Abdo:2010ex,Ackermann:2011wa,Ackermann:2013yva,Ackermann:2015zua, Charles:2016pgz, Fermi-LAT:2016uux}. At higher mass ranges, ground-based Cherenkov telescopes have advantages over observations with satellite detectors. Since most of those ground-based $\gamma$-ray facilities are pointing telescopes, 
upper limits on DM annihilation cross-sections are obtained by observations on a few well-selected dSphs. Almost the same level of upper limits is obtained by observations with different facilities~\cite{Aleksic:2011jx, Aleksic:2013xea,Doro:2017dqn,Ahnen:2017pqx,Aharonian:2007km,Aharonian:2008dm,Abramowski:2010aa, Aleksic:2013xea,Abramowski:2014tra, Acciari:2010ab,Aliu:2012ga,Archambault:2017wyh,Albert:2017vtb,Essig:2010em,Ahnen:2016qkx}.
In the very near future, the Cherenkov Telescope Array (CTA) starts its operations and is expected to improve the sensitivity to probe DM annihilation cross-sections by about one order of magnitude~\cite{Acharya:2017ttl}.

The designed angular resolution of CTA for $\gamma$-rays around 1TeV is ${\cal O}(0.05^\circ)$ degree, which is finer than the typical spatial extension of dSphs hence the consideration of the DM density profile shape becomes crucial. This has been pointed out in earlier works (e.g.~\cite{Charbonnier:2011ft,Ambrogi:2018skq}). In the latest analyses with atmospheric Cherenkov telescopes, spatial extensions of DM for dSphs are taken into account~\cite{Ahnen:2017pqx,Archambault:2017wyh} and tend to give upper limits milder than those assuming point sources. 
However, DM density distributions in the dSphs are still under discussion (see Appendix of Ref.~\cite{Charbonnier:2011ft} or Ref.~\cite{Bonnivard:2015vua} for examples). Different models for DM distributions lead to the divergence of derived upper limits. 
\begin{figure*}[th!!]
\begin{center}
\begin{tabular}{c}
\begin{minipage}{0.33\hsize}
\begin{center}
\includegraphics[width=5.5cm]{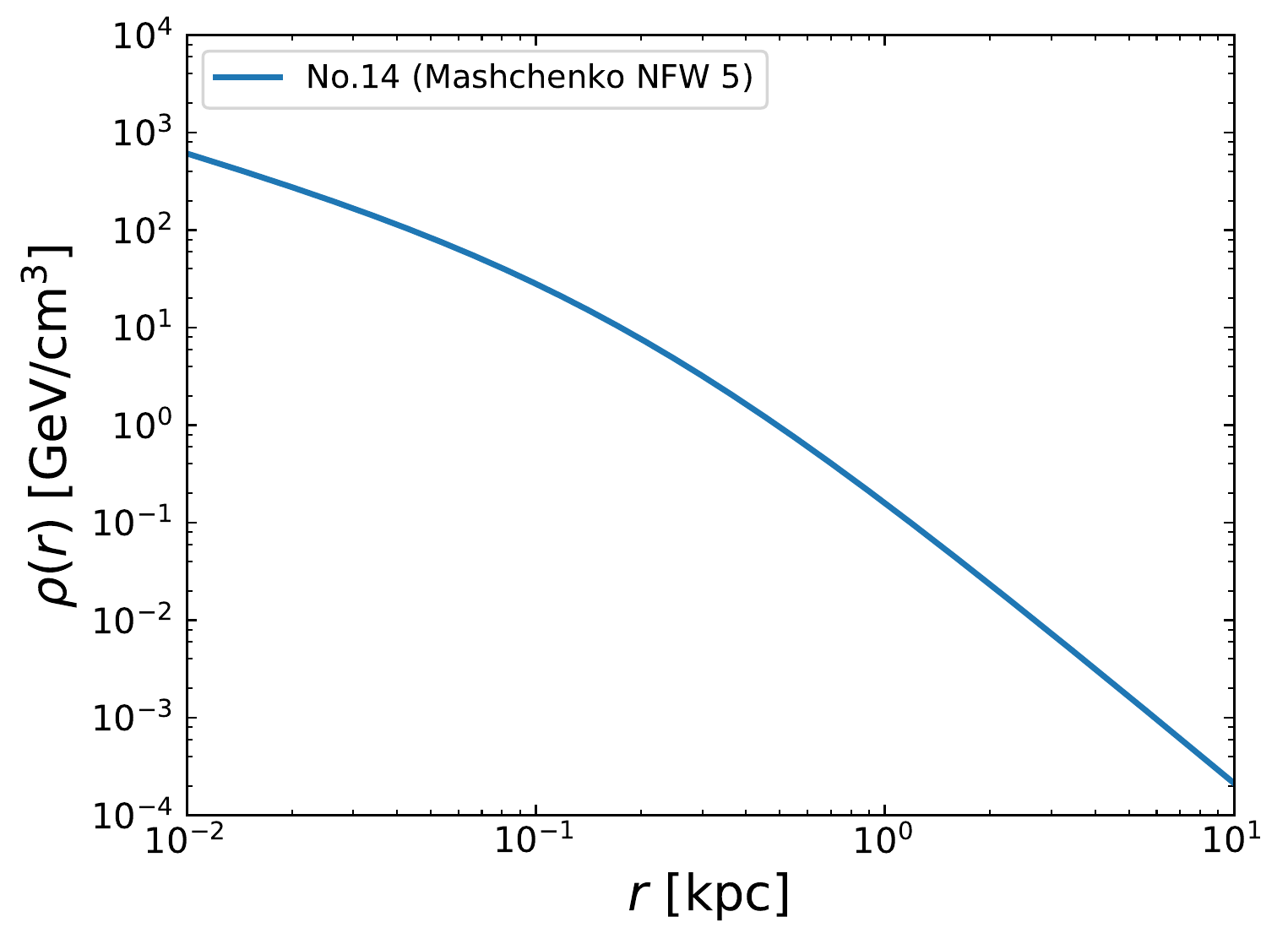} \hspace{0.8cm}
\end{center}
\end{minipage}
\begin{minipage}{0.33\hsize}
\begin{center}
\includegraphics[width=5.5cm]{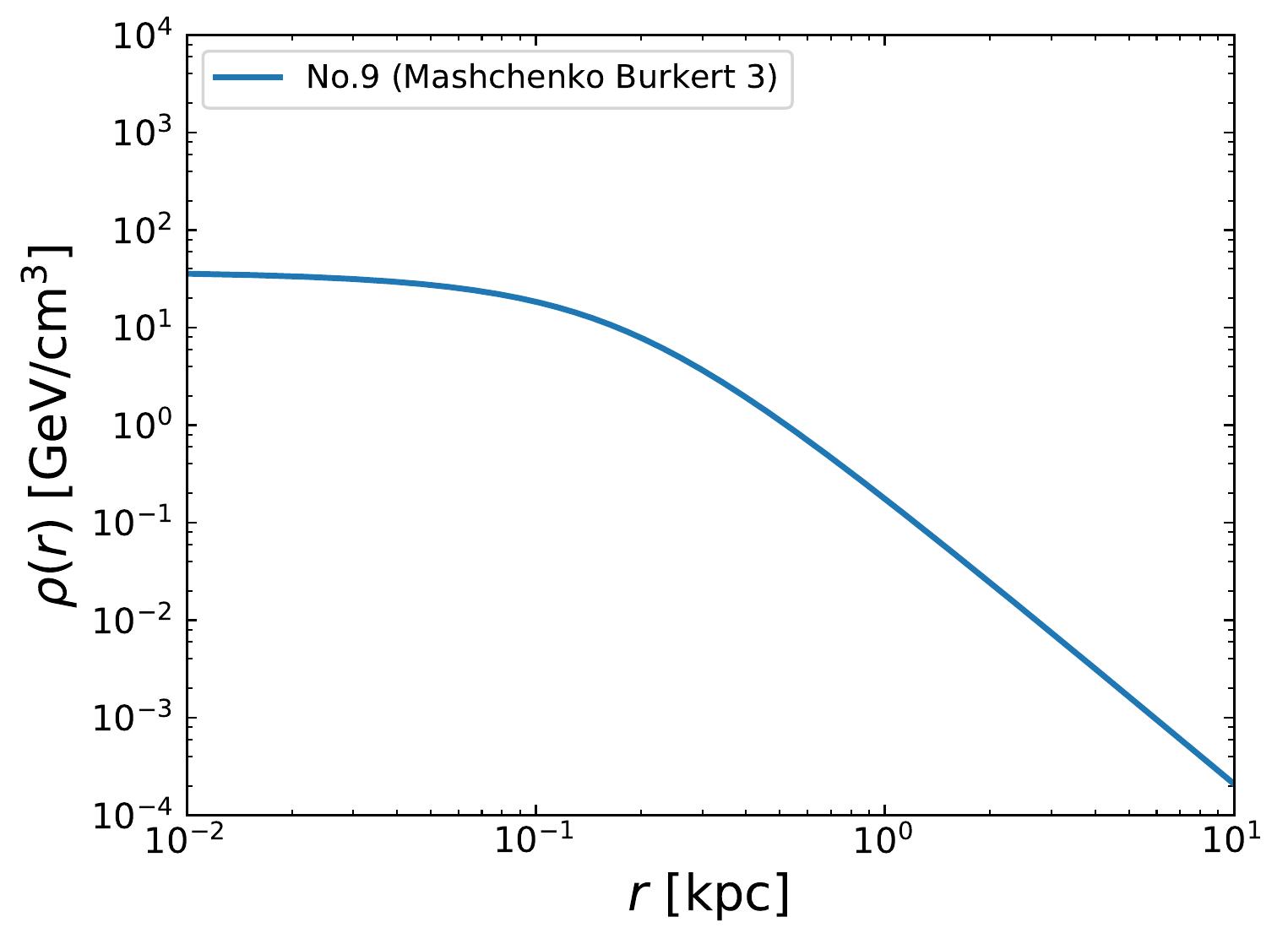} \hspace{0.8cm}
\end{center}
\end{minipage} 
\begin{minipage}{0.33\hsize}
\begin{center}
\includegraphics[width=5.5cm]{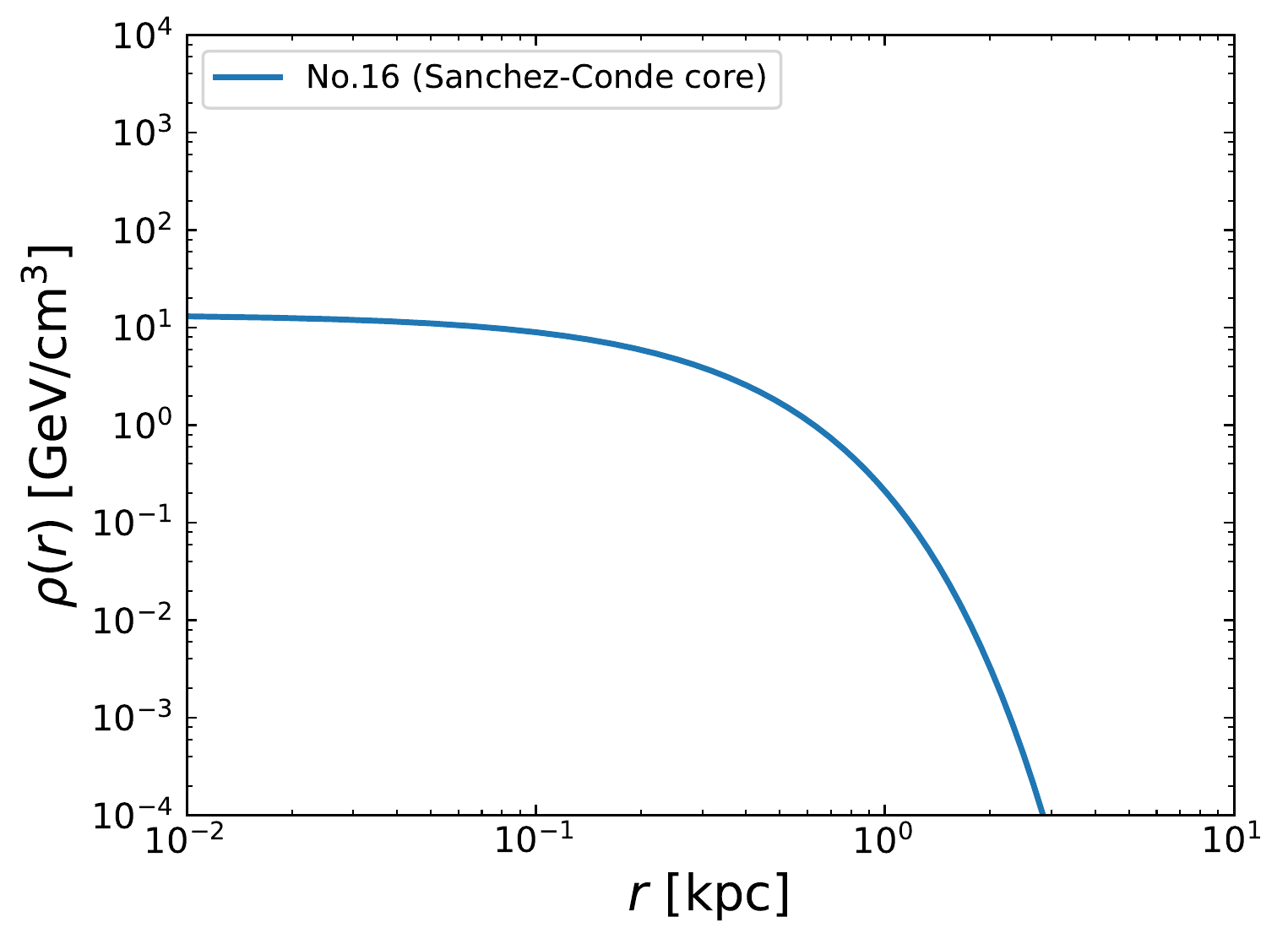} \hspace{0.8cm}
\end{center}
\end{minipage} 
\end{tabular}
\caption{Examples of DM density profiles in our analyses. ${\it Left}$: NFW profile of model 5 in \cite{Mashchenko:2005bj}. ${\it Center}$: Burkert profile of model 3 in \cite{Mashchenko:2005bj}. ${\it Right}$: power-law of (index 0) + cutoff profile in \cite{SanchezConde:2007te}. The horizontal axis represents the distance measured from the center of the dSph. Numbers in legends correspond to those in Table~\ref{tabref}.}
\label{fig:rdependence}
\end{center}
\end{figure*}

In this paper, we examine accessible parameter regions of the DM annihilation cross-section with CTA, probing different extended DM density distributions in dSphs. We sample DM density profiles of the Draco dSph as examples. Draco is one of the well-known classical dSph galaxies. So far, its several profiles
have been provided for it in the literature~\cite{Acciari:2010ab,Strigari:2006rd,Geringer-Sameth:2014yza,Lokas:2001mf,Lokas:2004sw,Mashchenko:2005bj,SanchezConde:2007te}. 
We consider the observation of dSphs with CTA and analyze simulated data using ${\tt ctools}$~ \cite{Knodlseder:2016nnv}.
The sensitivity calculations for DM annihilation cross-sections are conducted with16 different profiles and compared to give a quantitative estimate of uncertainties in the searches towards dSphs. The dependence on the distances to dSphs is also investigated.

The structure of this paper is as follows. Sec.~\ref{sec:methods} explains our methods. In Sec.~\ref{sec:results} we show a comparison of the sensitivity for annihilation cross-sections obtained with various profiles and distances. Sec.~\ref{sec:discussion} is devoted to discussions. We summarise in Sec.\ref{sec:conclusion}.

\section{Methods}
\label{sec:methods}

\begin{figure*}[th!!]
\begin{center}
\begin{tabular}{c}
\begin{minipage}{0.33\hsize}
\begin{center}
\includegraphics[width=5.5cm]{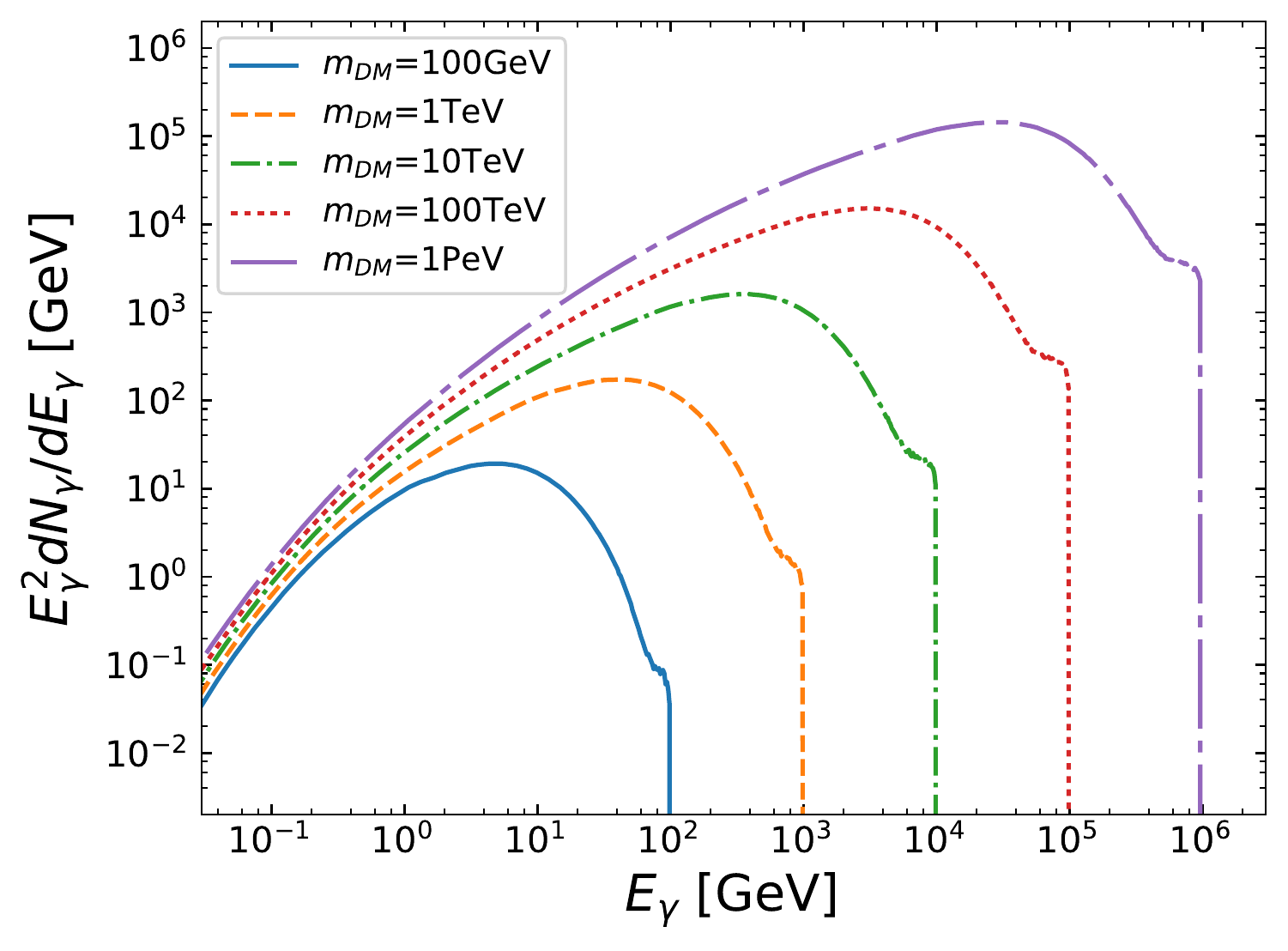} \hspace{0.8cm}
\end{center}
\end{minipage}
\begin{minipage}{0.33\hsize}
\begin{center}
\includegraphics[width=5.5cm]{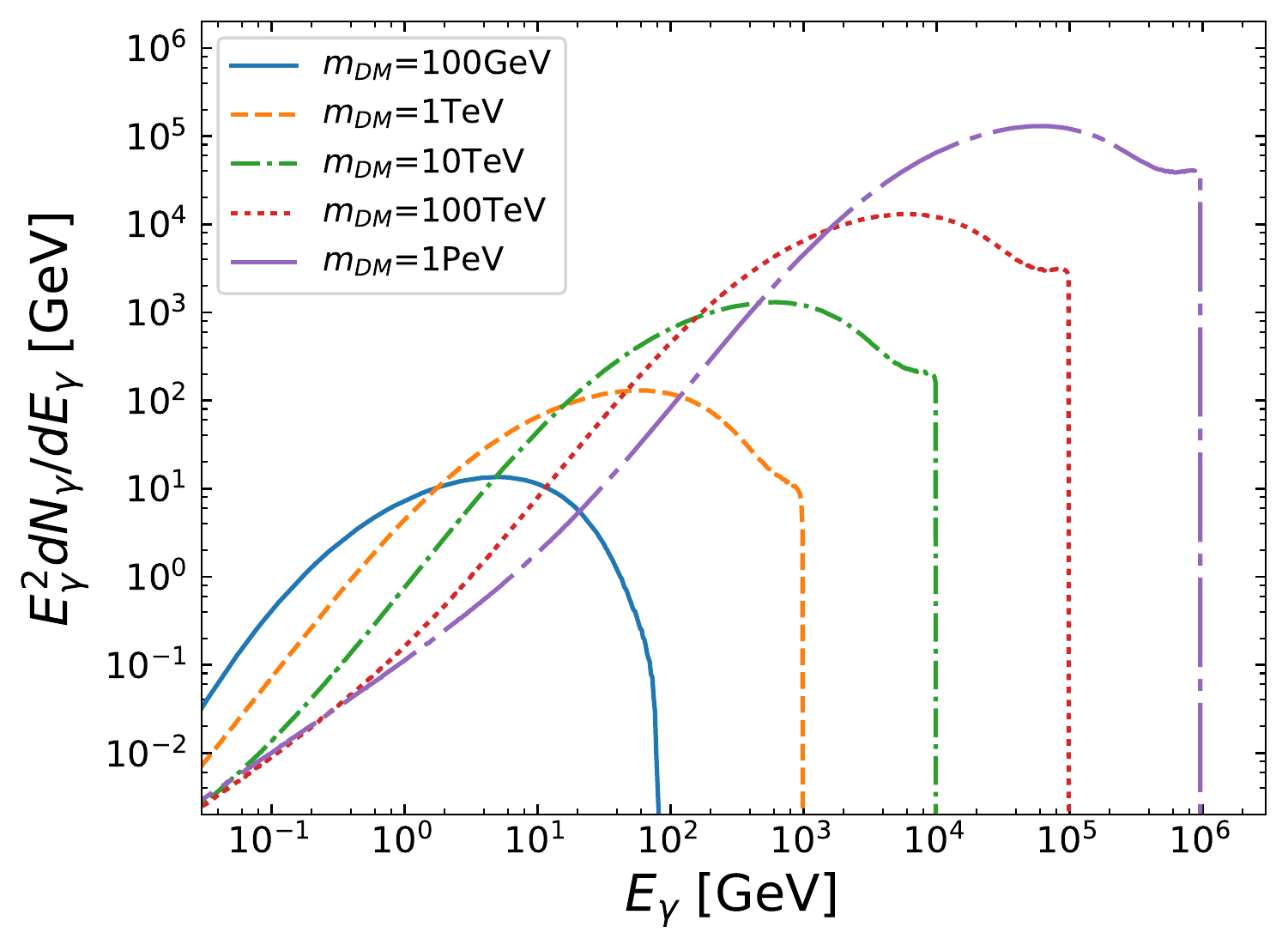} \hspace{0.8cm}
\end{center}
\end{minipage} 
\begin{minipage}{0.33\hsize}
\begin{center}
\includegraphics[width=5.5cm]{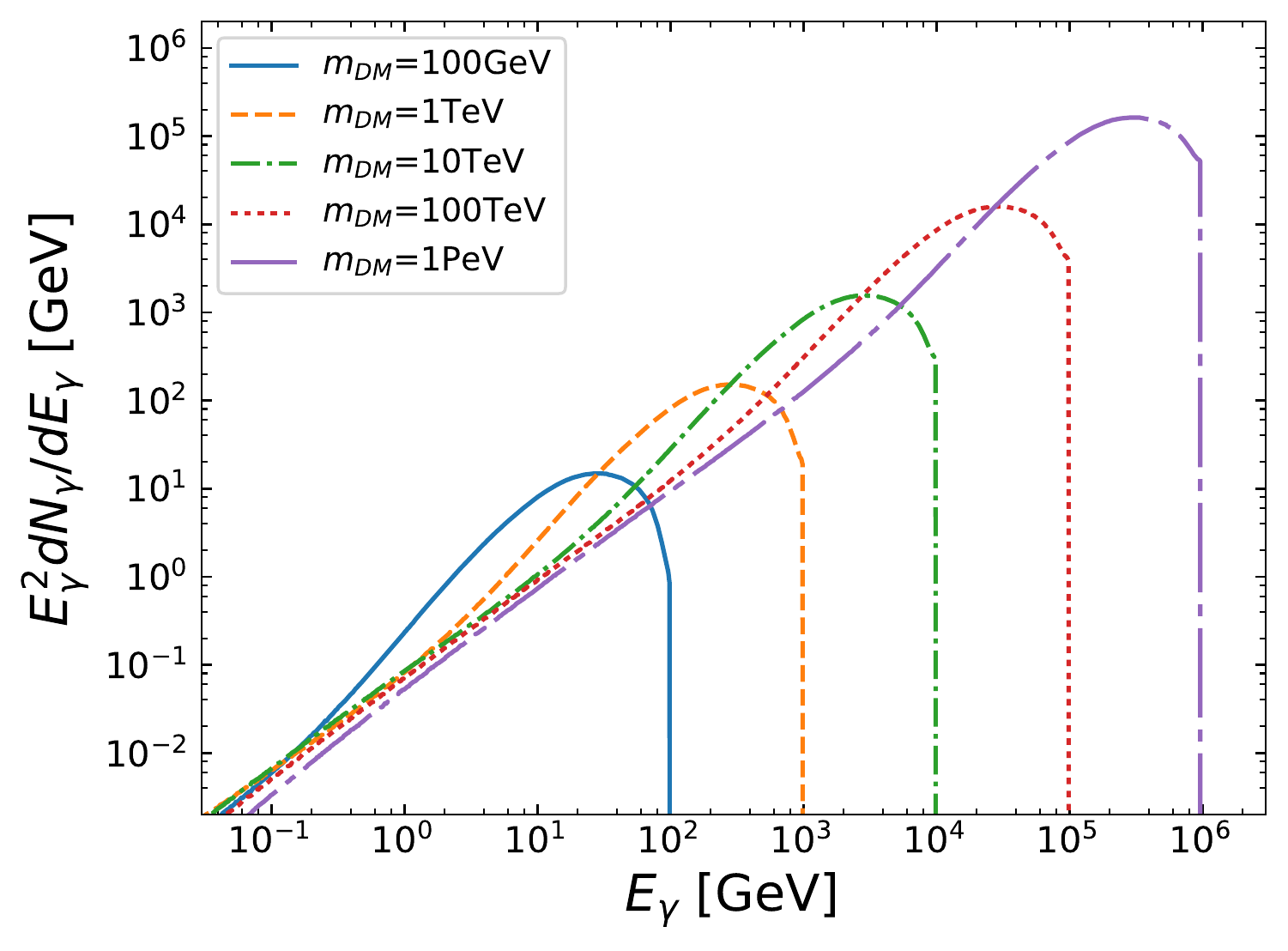} \hspace{0.8cm}
\end{center}
\end{minipage} 
\end{tabular}
\caption{$\gamma$-ray spectrum for DM of mass $m_\mathrm{DM}=$100GeV, 1TeV, 10TeV, 100TeV and 1PeV annihilating into $\bar{b}b$ (${\it left}$), $W^+W^-$ (${\it center}$) and $\tau^+\tau^-$ (${\it right}$). Spectrum of $m_\mathrm{DM}=100$GeV annihilating into $W^+W^-$ is shown for comparison and not used in our analyses.}
\label{srcspectrum}
\end{center}
\end{figure*}

\subsection{Dark matter density profiles of the source}
\label{ssec:profile}
A point source is the simplest model for a target dSph 
when the angular size of the target is small enough compared to the angular resolution of observational facilities. Future ground-based atmospheric Cherenkov telescopes can resolve typical dSphs, so they are to be treated as extended sources. Profiles of dSphs are sampled to investigate how their spatial extension affects the accessible region of the DM annihilation cross-section. Draco dSph is taken as the example, and we limit our analyses to spherical profiles for simplicity. Three types of DM density profiles are considered in this work:
\begin{enumerate}
\item generalized NFW profile \cite{Hernquist:1990be,Zhao:1995cp}:
 \begin{equation}
\rho(r)=\rho_s\left(\frac{r}{r_s}\right)^{-\gamma}\left(1+\left(\frac{r}{r_s}\right)^\alpha\right)^{-\frac{\beta-\gamma}\alpha},
\end{equation}
where ($\alpha$, $\beta$, $\gamma$)=(1, 3, 1) corresponds to the original NFW profile in \cite{Navarro:1996gj}.
\item Burkert profile \cite{Burkert:1995yz} :
\begin{equation}
\rho(r)=\rho_s\left(1+\frac{r}{r_s}\right)^{-1}\left(1+\left(\frac{r}{r_s}\right)^2\right)^{-1},
\end{equation}
\item power law (PL) profile with an exponential cutoff:
\begin{equation}
\rho(r)=\rho_s\left(\frac{r}{r_s}\right)^{-\alpha}\exp\left[-\frac{r}{r_s}\right].
\end{equation}
\end{enumerate}
$\rho_s$ is the normalization of the DM density, and $r_s$ is the scale radius of the profile measuring the distance $r$ from the center of the target. More detailed profiles such as non-spherical cases or profiles with substructures are discussed in Ref.~\cite{Bonnivard:2014kza,Bonnivard:2015pia,Hayashi:2016kcy,Hutten:2016jko}. Table~\ref{tabref} summarises our reference profiles with explicit expressions of each profile, profile type (corresponding to Eqs.~(3), (4), and(5)), J-factor integrated to solid angle of $0.5$ degrees ($J_{<0.5^\circ}$), J-factor integrated to $4.0^\circ\times4.0^\circ$  region ($J_\mathrm{tot}$) which corresponds to the size of the region of interest (RoI) in our analyses, and distance from the Earth. We also assign identification numbers in the first column in Table~\ref{tabref} for convenience. Note that the truncation radius for the profiles is not introduced in our analyses. The truncation radius is usually determined by the location of the outermost member star or the virial radius of the DM halo. If we take the former for the truncation radius, then it corresponds to $\theta=1.3^\circ$ for Draco~\cite{Geringer-Sameth:2014yza}. On the other hand, the virial radius is highly model-dependent. Actual radial extension of the dSphs is still under discussion. We chose our RoI to cover the outermost member star, avoiding the introduction of an additional model parameter. The J-factor integrated to 1.3$^\circ$ and $J_{\rm tot}$ defined as J-factors in our RoI differ by at most 10\%. Templates of the J-factor centered on the target are generated adopting the median value of the parameters for each profile provided in references. The spatial resolution of our template is 0.01$^\circ$. In practice, we produce templates larger than the RoI, then use parts corresponding to the RoI. $J_{<0.5^\circ}$ values in Table~\ref{tabref} are shown just to make a comparison with previous works easier and are not used in our analyses. 

\subsection{Spectrum of the DM annihilation at the source}
\label{ssec:spectrum}
Three channels are considered as final states, $\bar{b}b$, $W^+W^-$ and $\tau^+\tau^-$. Those are representatives of DM annihilations into quarks, weak bosons, and leptons. The maximum mass of the DM particle in our calculation is set to $m_\mathrm{DM}$ = 1PeV, while the minimum to $m_\mathrm{DM}=$ 25GeV for lepton and quark channels, and to $m_\mathrm{DM}=$ 160GeV for the weak boson channel. At lower energies, contributions from residual cosmic-rays are significant. We set our minimum mass so that to avoid these contaminations.
 The spectra of each annihilation channel are calculated with
${\tt
  pythia8.2}$~\cite{Sjostrand:2014zea,Sjostrand:2006za,Sjostrand:2007gs}.
Figure~\ref{srcspectrum} shows examples of spectra from
$m_\mathrm{DM}=$100GeV to 1PeV. 
The spectra shown in Figure~\ref{srcspectrum} include  final state
radiations like Bremsstrahlung of charged leptons, which are
electroweak corrections different from interactions with external
fields. We consider contributions from secondary $\gamma$-rays
produced during propagations of charged leptons to be
negligible~\cite{Belikov:2009cx,Cirelli:2009vg,Profumo:2009uf,Blanco:2017sbc,Bartels:2017dpb}. The
treatment of the secondary $\gamma$-rays and the spectra in our
calculation would be consistent with those available in
Ref.~\cite{Cirelli:2010xx} which are computed by the old version {\tt pythia8.1} and widely used in
$\gamma$-ray searches of dark matter.

 The differences in the gamma-ray spectra between $W^+W^-$ (or $\bar{b}b$) and
$\tau^+\tau^-$ modes come from differences of the particle multiplicity
among those modes. $\gamma$-rays are produced mainly by decaying neutral
pions, and partly by other decaying mesons. In the $W^+W^-$ or $\bar{b}b$ modes,
emitted quark-pairs immediately fragment into a lot of mesons and
baryons, which are dominant modes. The number of the multiplicity into
pions would be approximately $\sim$ 30 for the center-of-momentum energy
being $\sqrt{s}$ = O(1) TeV. In this case, the $\gamma-ray$ spectrum
becomes broader with its mean energy being lower.  On the other hand,
in the $\tau^+\tau^-$ mode, the number of the multiplicity into neutral pions
is much smaller (a few in $\sqrt{s}$ = O(1)TeV). In this latter case,
the energy of $\gamma$-rays tends to be higher, which gives the steeper
spectrum than that of the $W^+W^-$ or $\bar{b}b$ emission mode.

\begingroup
\scalefont{0.75}
\begin{table*}[h!!t!]
\caption{DM density profiles for dSphs used in our analysis. We assign numbers in the first column for convenience. We adopt the center value for the parameters in each case.}
\begin{tabular}{ccccccc}\hline
No.&Reference&expression&type&log$_{10}$J$_{<0.5^\circ}$&log$_{10}$J$_\mathrm{tot}$&distance[kpc]\\ \hline
1&Acciari et al. \cite{Acciari:2010ab}&$\left(\frac{1.7{\rm GeV}}{{\rm cm}^3}\right)\left(\frac{r}{0.79\mathrm{kpc}}\right)^{-1}\left(1+\frac{r}{0.79\mathrm{kpc}}\right)^{-2}$&NFW&18.40&18.45&80 \rule[0mm]{0mm}{5mm}\\ \hline
2&Geringer-Sameth et al.\cite{Geringer-Sameth:2014yza}&$\left(\frac{0.69{\rm GeV}}{{\rm cm}^3}\right)\left(\frac{r}{3.7{\rm kpc}}\right)^{-0.71}\left(1+\left(\frac{r}{3.7{\rm kpc}}\right)^{2.01}\right)^{-2.80}$&generalized NFW&19.00&19.44&76\rule[0mm]{0mm}{5mm}\\ \hline
3&Lokas\cite{Lokas:2001mf}&$\left(\frac{16.3{\rm GeV}}{{\rm cm}^3}\right)\left(1+\frac{r}{0.67{\rm kpc}}\right)^{-3}$&generalized NFW&19.08&19.29& \rule[0mm]{0mm}{5mm}\\
4&&$\left(\frac{1.23{\rm GeV}}{{\rm cm}^3}\right)\left(\frac{r}{1.30{\rm kpc}}\right)^{-1}\left(1+\frac{r}{1.30{\rm kpc}}\right)^{-2}$&NFW&18.80&18.91&72\rule[0mm]{0mm}{5mm}\\
5&&$\left(\frac{0.18{\rm GeV}}{{\rm cm}^3}\right)\left(\frac{r}{1.99{\rm kpc}}\right)^{-1.5}\left(1+\frac{r}{1.99{\rm kpc}}\right)^{-1.5}$&generalized NFW&18.88&18.90& \rule[0mm]{0mm}{5mm}\\ \hline
6&Lokas et al. \cite{Lokas:2004sw}&$\left(\frac{5.9{\rm GeV}}{{\rm cm}^3}\right)\left(\frac{r}{0.32{\rm kpc}}\right)^{-1}\exp\left[-\frac{r}{0.32{\rm kpc}}\right]$&PL + cutoff&18.53&18.53&80 \rule[0mm]{0mm}{5mm} \\ \hline
7&Mashchenko et al. \cite{Mashchenko:2005bj}&$\left(\frac{4.76{\rm GeV}}{{\rm cm}^3}\right)\left(1+\frac{r}{1.41{\rm kpc}}\right)^{-1}\left(1+\left(\frac{r}{1.41{\rm kpc}}\right)^2\right)^{-1}$&Burkert&19.08&19.56&\rule[0mm]{0mm}{5mm} \\
8&&$\left(\frac{13.4{\rm GeV}}{{\rm cm}^3}\right)\left(1+\frac{r}{0.35{\rm kpc}}\right)^{-1}\left(1+\left(\frac{r}{0.35{\rm kpc}}\right)^2\right)^{-1}$&Burkert&18.65&18.70&\rule[0mm]{0mm}{5mm}\\
9&&$\left(\frac{37.8{\rm GeV}}{{\rm cm}^3}\right)\left(1+\frac{r}{0.18{\rm kpc}}\right)^{-1}\left(1+\left(\frac{r}{0.18{\rm kpc}}\right)^2\right)^{-1}$&Burkert&18.69&18.70&\rule[0mm]{0mm}{5mm}\\
10&&$\left(\frac{0.60{\rm GeV}}{{\rm cm}^3}\right)\left(\frac{r}{2.82\mathrm{kpc}}\right)^{-1}\left(1+\frac{r}{2.82\mathrm{kpc}}\right)^{-2}$&NFW&18.95&19.15&82\rule[0mm]{0mm}{5mm}\\
11&&$\left(\frac{1.70{\rm GeV}}{{\rm cm}^3}\right)\left(\frac{r}{1.00\mathrm{kpc}}\right)^{-1}\left(1+\frac{r}{1.00\mathrm{kpc}}\right)^{-2}$&NFW&18.67&18.73&\rule[0mm]{0mm}{5mm}\\
12&&$\left(\frac{4.76{\rm GeV}}{{\rm cm}^3}\right)\left(\frac{r}{0.50\mathrm{kpc}}\right)^{-1}\left(1+\frac{r}{0.50\mathrm{kpc}}\right)^{-2}$&NFW&18.70&18.72&\rule[0mm]{0mm}{5mm}\\
13&&$\left(\frac{13.4{\rm GeV}}{{\rm cm}^3}\right)\left(\frac{r}{0.25\mathrm{kpc}}\right)^{-1}\left(1+\frac{r}{0.25\mathrm{kpc}}\right)^{-2}$&NFW&18.70&18.70&\rule[0mm]{0mm}{5mm}\\
14&&$\left(\frac{37.8{\rm GeV}}{{\rm cm}^3}\right)\left(\frac{r}{0.18\mathrm{kpc}}\right)^{-1}\left(1+\frac{r}{0.18\mathrm{kpc}}\right)^{-2}$&NFW&19.15&19.15&\rule[0mm]{0mm}{5mm}\\ \hline
15&Sanchez-Conde et al. \cite{SanchezConde:2007te}&$\left(\frac{0.95{\rm GeV}}{{\rm cm}^3}\right)\left(\frac{r}{1.19{\rm kpc}}\right)^{-1}\exp\left[-\frac{r}{1.19{\rm kpc}}\right]$&PL + cutoff&18.58&18.69&80\rule[0mm]{0mm}{5mm}\\
16&&$\left(\frac{12.7{\rm GeV}}{{\rm cm}^3}\right)\exp\left[-\frac{r}{0.24{\rm kpc}}\right]$&PL + cutoff&18.56&18.58&\rule[0mm]{0mm}{5mm} \\ \hline
\end{tabular}
\label{tabref}
\end{table*}
\endgroup

\subsection{General procedures of our analysis}
\label{ssec:genral_p}
 The procedure for sensitivity calculations is as follows. The software package ${\tt ctools}$~\cite{Knodlseder:2016nnv} is used for the analysis. First, we simulate events assuming a 500-hour observation. The instrumental response function (IRF) ${\tt prod3b}$ ~\cite{Cumani:2017aca}, the latest publicly available version of the CTA IRF package, is used. Assuming the northern CTA site (La Palma), we select the IRF optimized for the long-time observation at a zenith angle of 20 degrees.
 In the event generations, no $\gamma$-ray sources are included. Only residual charged cosmic rays as background events are simulated. 
After the event generation, selections and binnings are performed in energy and space. We select a 4$^\circ\times4^\circ$ square region centered on the target. Spatial binning is 0.03$^\circ$. In energy, events from 0.03 TeV to 180 TeV are selected and binned with 5 bins per decade on a logarithmic scale. We conduct likelihood analyses with the binned data. Median upper limits on the $\gamma$-ray flux are defined as to decrease the likelihood corresponding to a 95\% confidence level. Throughout the procedure, we calculate with ${\it ctools}$ following the method in~\cite{Knodlseder:2016nnv}. The dependence between the $\gamma$-ray flux and annihilation cross-section is given in Eq.~\ref{eq:gammaflux}. 

\section{Results}
\label{sec:results}
\begin{figure*}[h!!!pt]
\begin{center}
\begin{tabular}{c}
\begin{minipage}{0.33\hsize}
\begin{center}
\includegraphics[width=6cm]{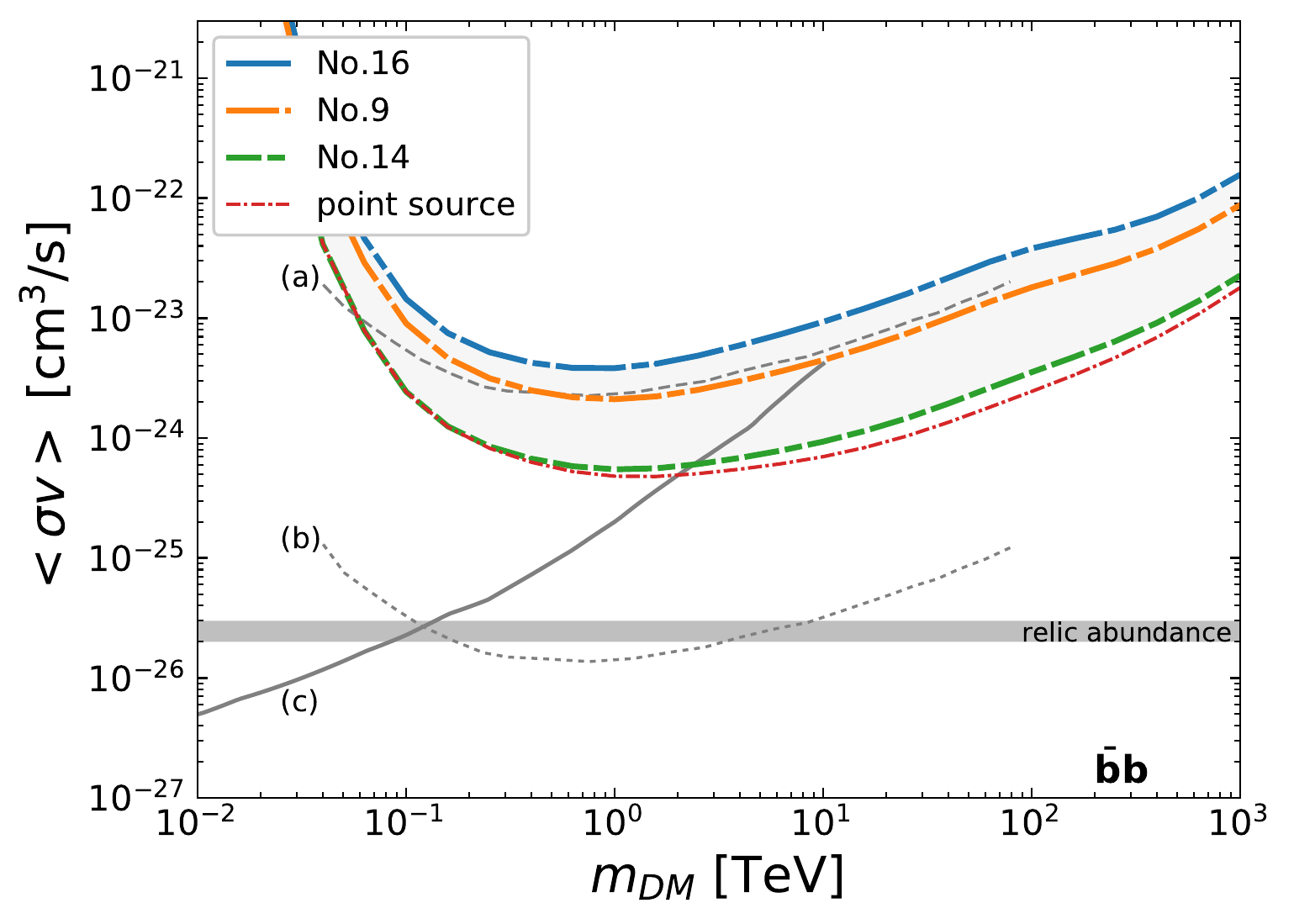} \hspace{0.8cm}
\end{center}
\end{minipage}
\begin{minipage}{0.33\hsize}
\begin{center}
\includegraphics[width=6cm]{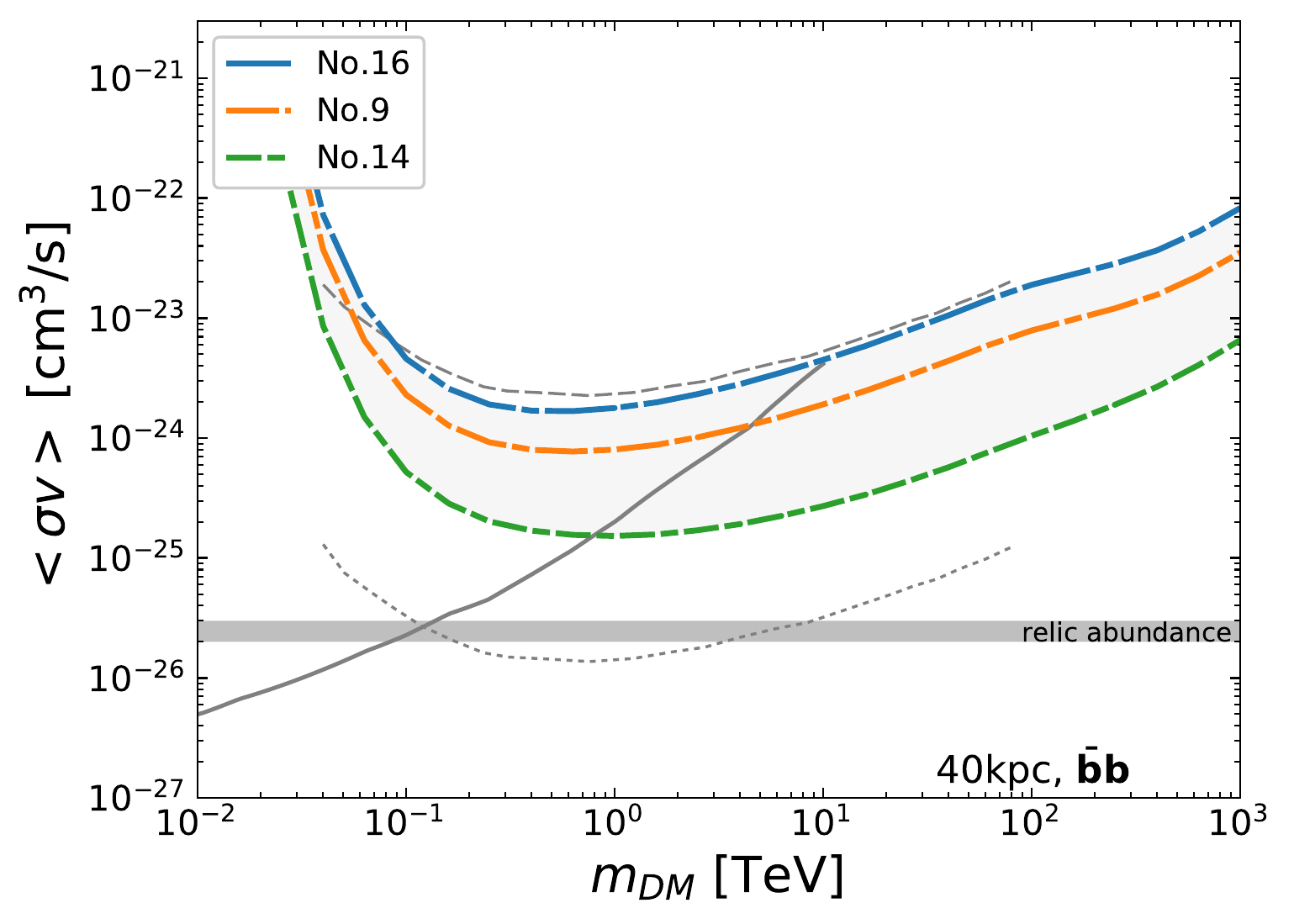} \hspace{0.8cm}
\end{center}
\end{minipage}
\begin{minipage}{0.33\hsize}
\begin{center}
\includegraphics[width=6cm]{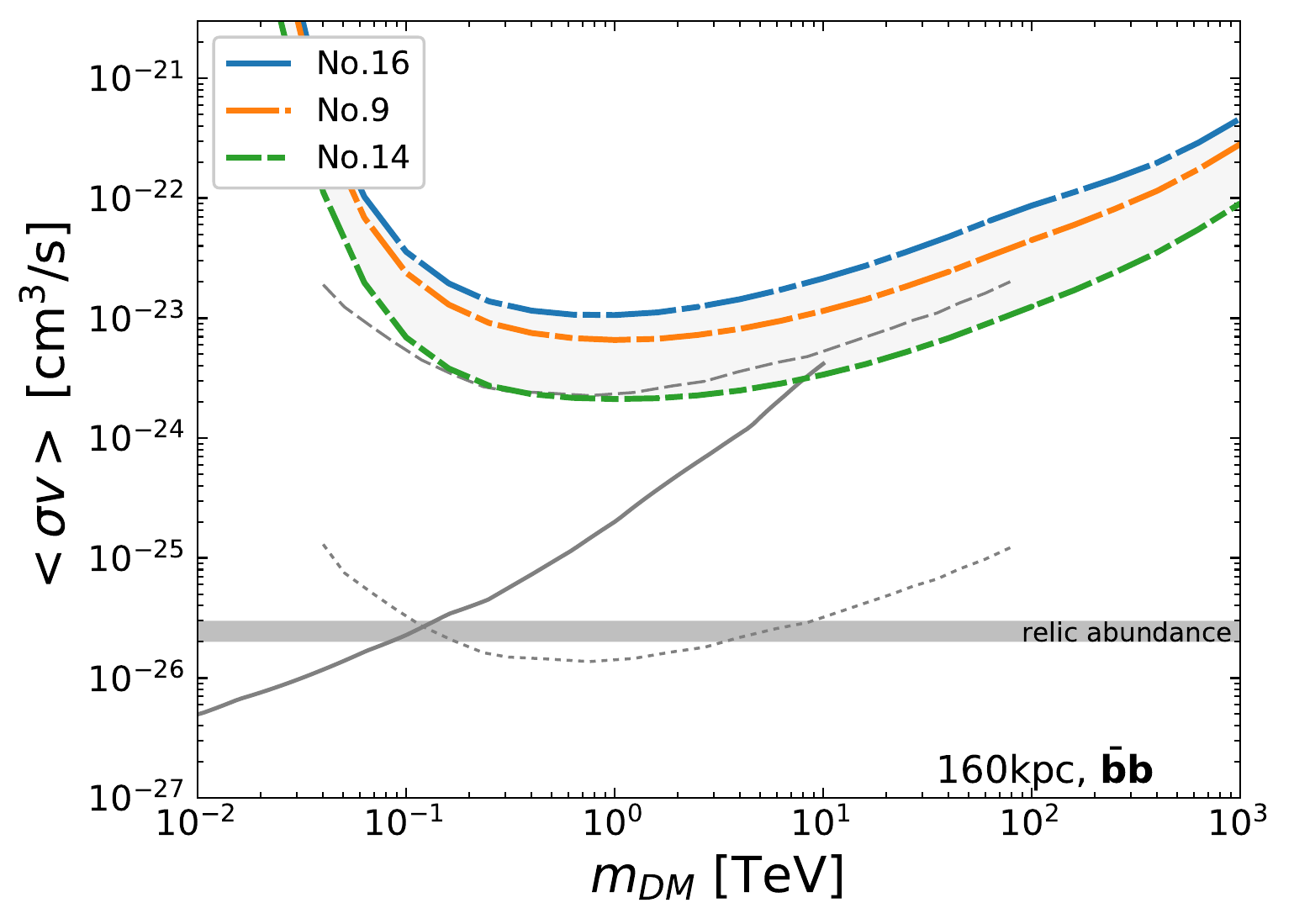} \hspace{0.8cm}
\end{center}
\end{minipage} \\
\begin{minipage}{0.33\hsize}
\begin{center}
\includegraphics[width=6cm]{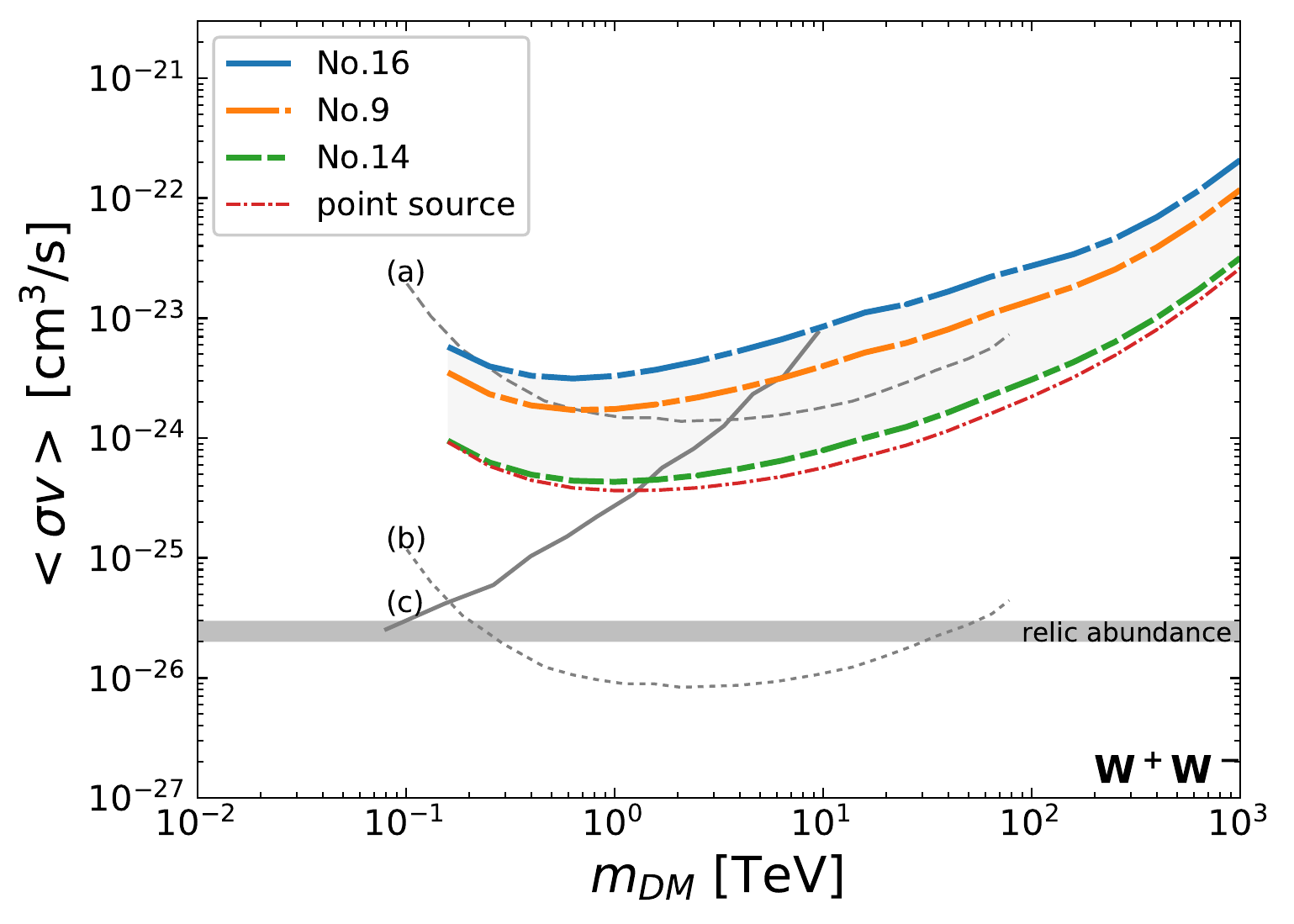} \hspace{0.8cm}
\end{center}
\end{minipage}
\begin{minipage}{0.33\hsize}
\begin{center}
\includegraphics[width=6cm]{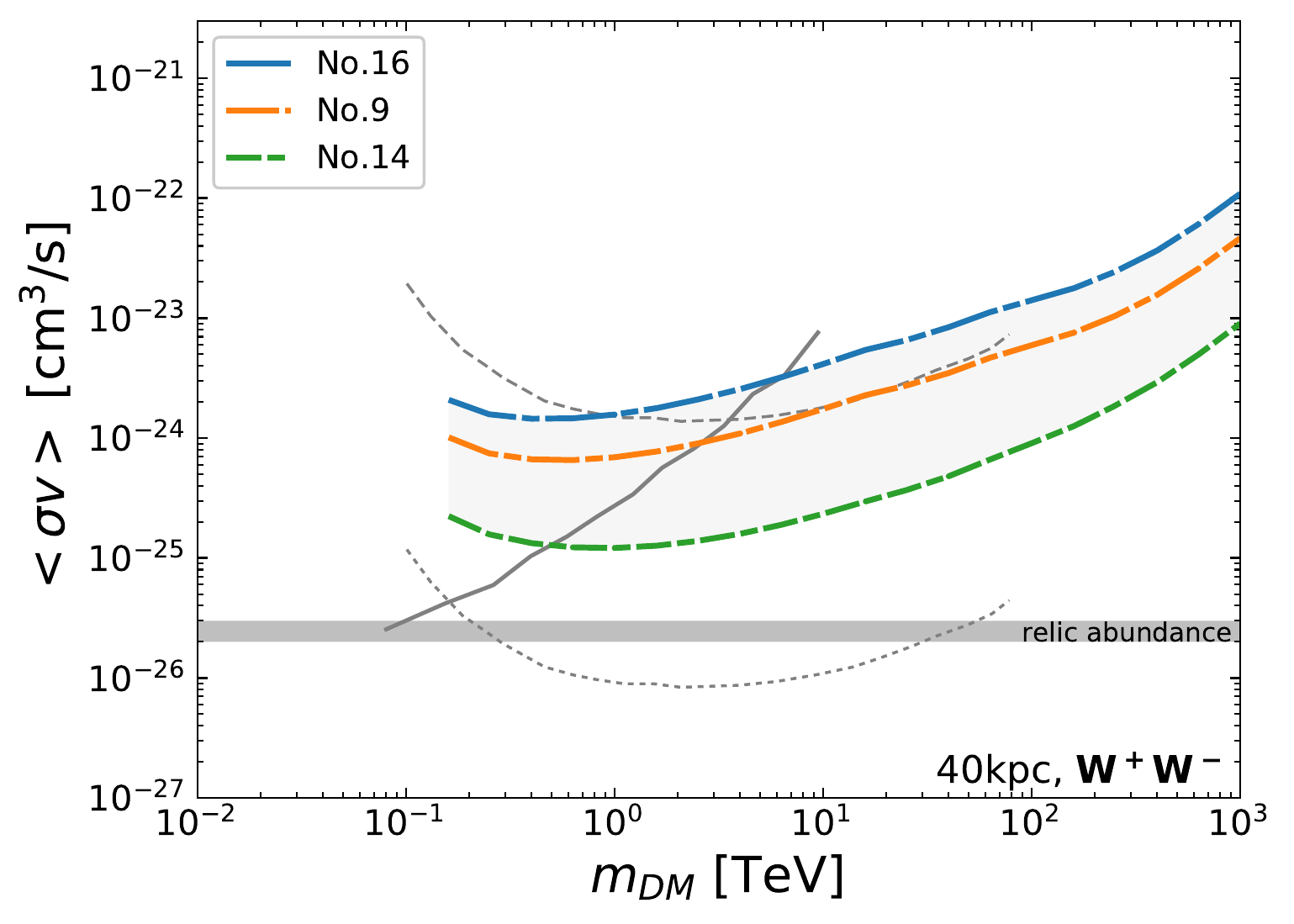} \hspace{0.8cm}
\end{center}
\end{minipage}
\begin{minipage}{0.33\hsize}
\begin{center}
\includegraphics[width=6cm]{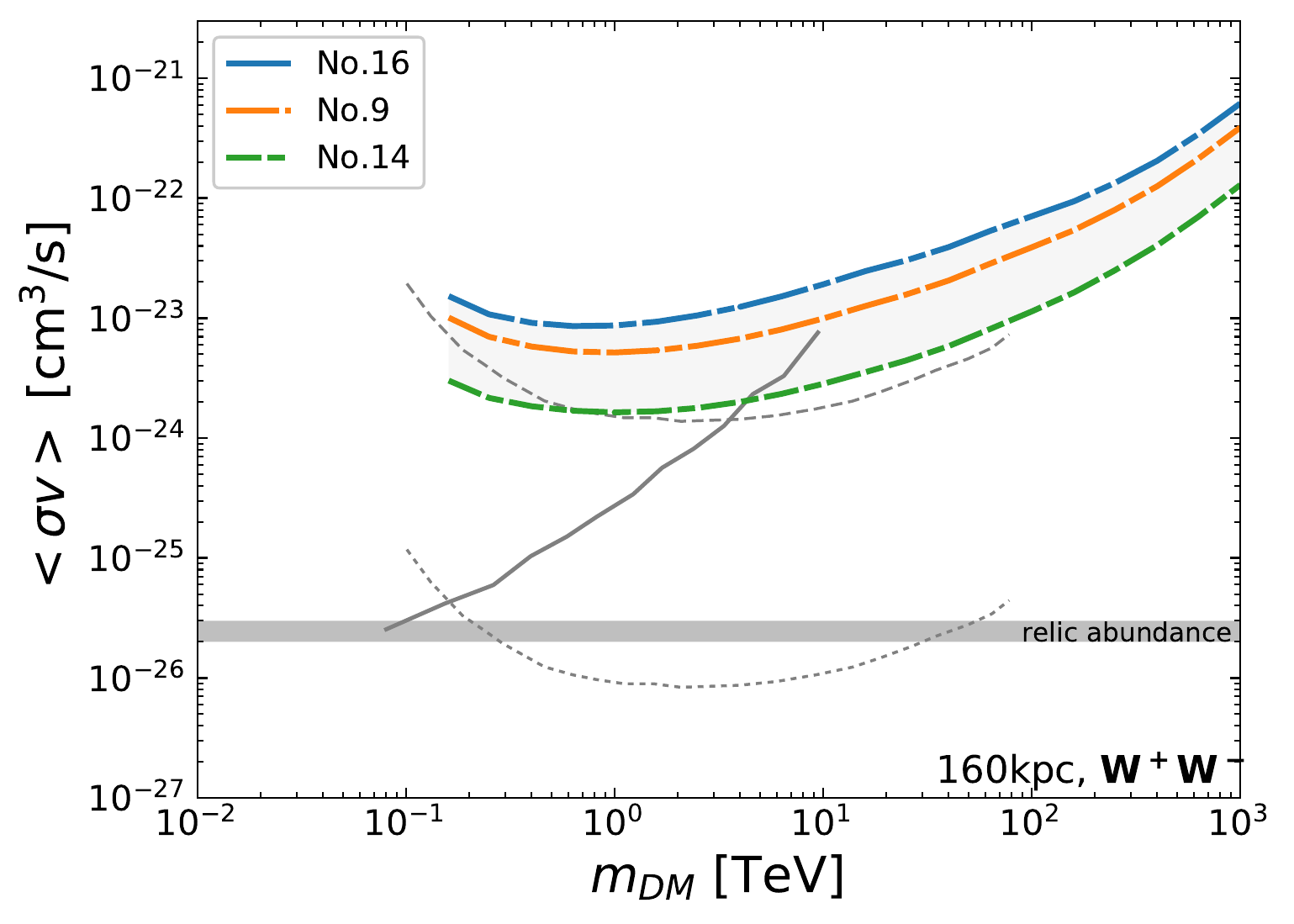} \hspace{0.8cm}
\end{center}
\end{minipage} \\
\begin{minipage}{0.33\hsize}
\begin{center}
\includegraphics[width=6cm]{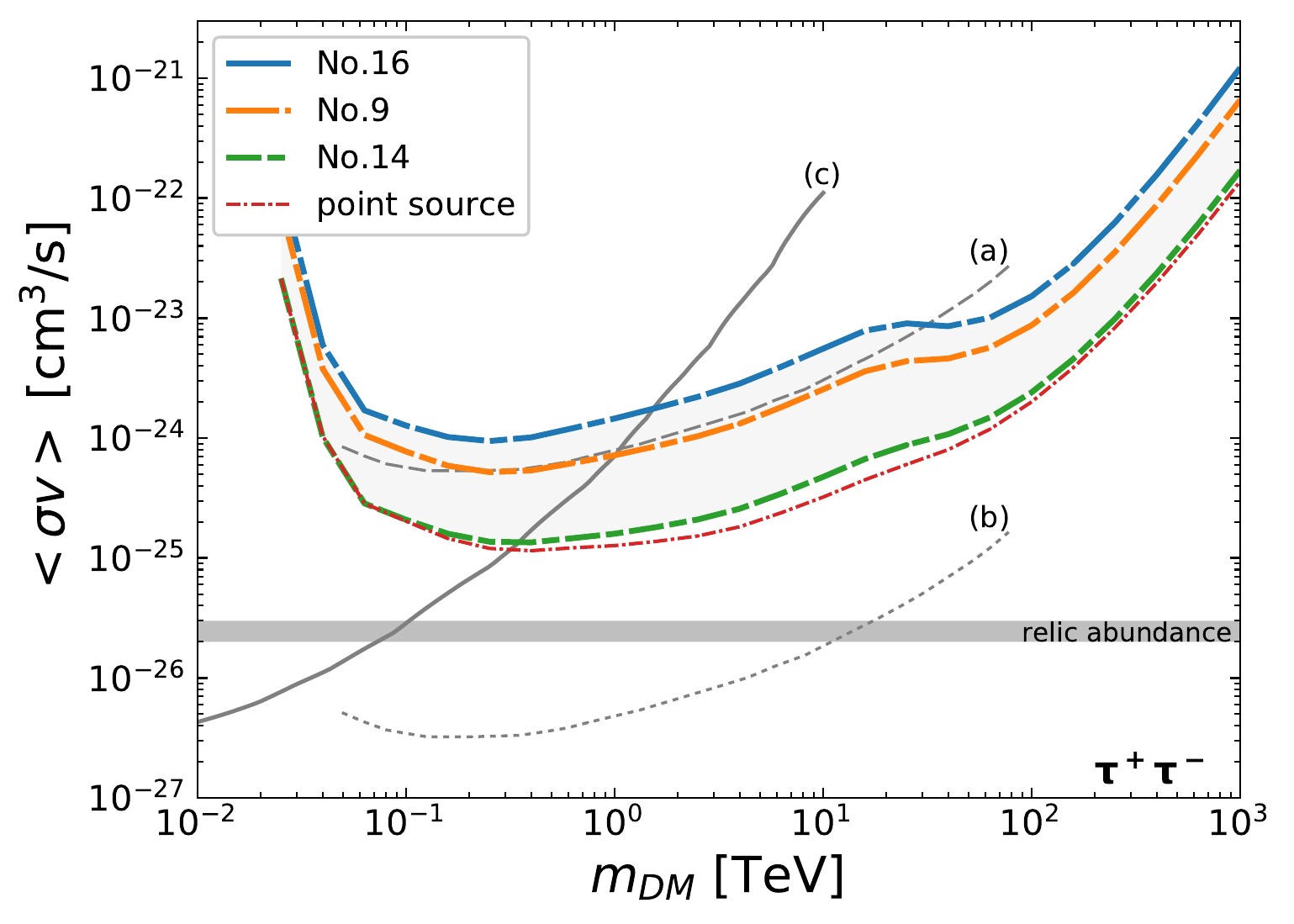} \hspace{0.8cm}
\end{center}
\end{minipage}
\begin{minipage}{0.33\hsize}
\begin{center}
\includegraphics[width=6cm]{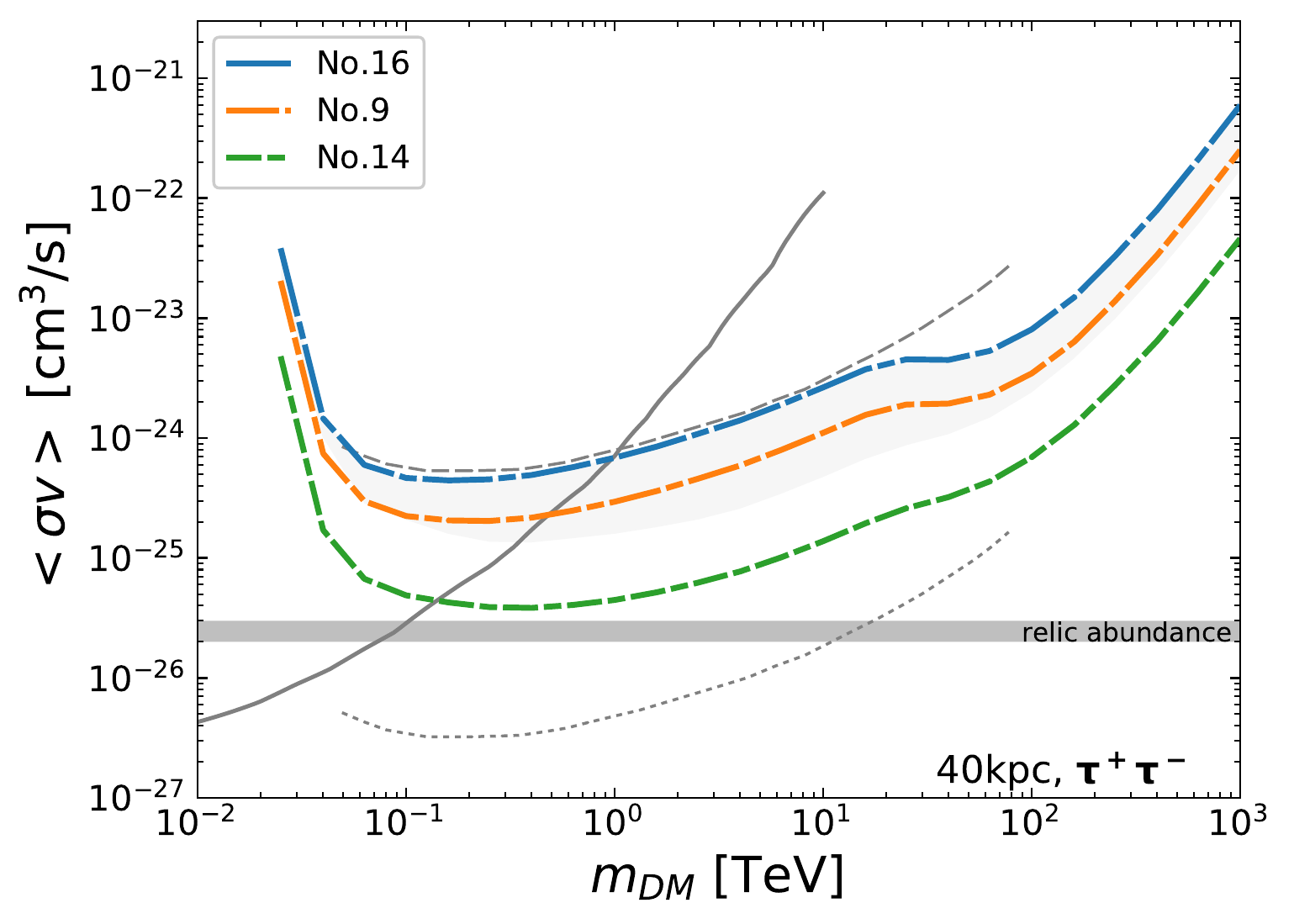} \hspace{0.8cm}
\end{center}
\end{minipage}
\begin{minipage}{0.33\hsize}
\begin{center}
\includegraphics[width=6cm]{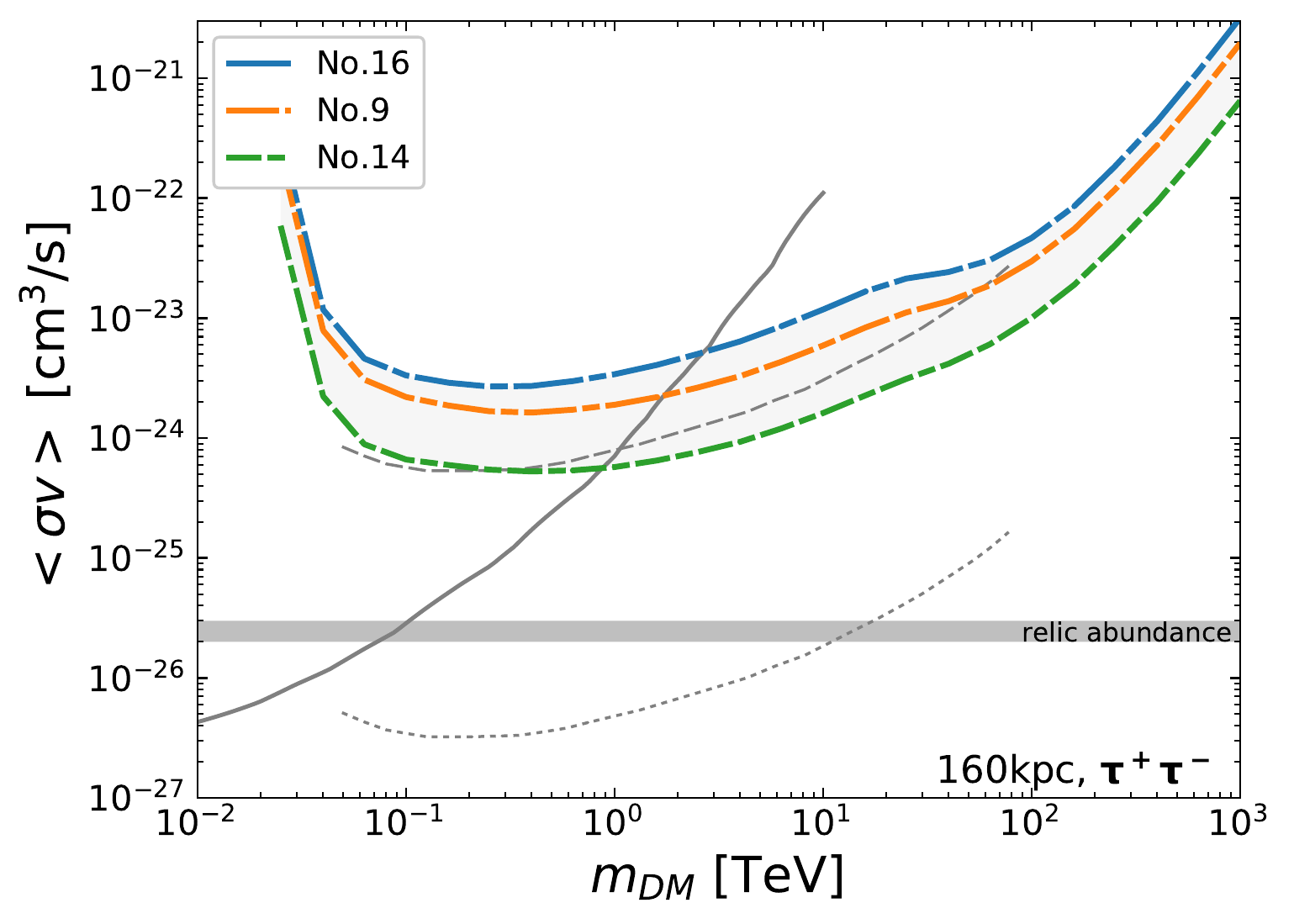} \hspace{0.8cm}
\end{center}
\end{minipage}
\end{tabular}
\caption{Upper limits on the DM annihilation cross-section assuming a 500-hour observation. In each panel, we show our results with dash-dotted lines using several dash length. For comparison, we put three lines: (a) The CTA sensitivity curve for a 500-hour observation of the Galactic halo with CTA~\cite{Acharya:2017ttl} assuming a Burkert profile (long-dashed line), (b) the same but assuming an Einasto profile (short-dashed line), and (c) an observational result of the current upper limit by {\it Fermi}-LAT using 25 dSphs with kinematically derived J-factors~~\cite{Ackermann:2015zua} (solid line). We also show a reference for the cross-section $\left<\sigma v\right>$ corresponding to the relic abundance in a horizontal band in the bottom part. Detailed calculations for the relic abundance is given in e.g. Ref.~\cite{Steigman:2012nb}. Panels in the left column show sensitivities assuming profiles in Table~\ref{tabref}. The distance for each profile is also shown in the same table. The severest case of $J_\mathrm{tot}=10^{19.15}$ (No. 14 in Table~\ref{tabref}, {\it left} panel in Figure~\ref{fig:rdependence}) is shown with a short dashed-dotted line and the weakest one of $J_\mathrm{tot}=10^{18.58}$ (No. 16, {\it right} pael in Figure~\ref{fig:rdependence}) with a long dash-dotted line. Middle dash-dotted lines correspond to the case of {\it center} panel in Figure~\ref{fig:rdependence} of $J_\mathrm{tot}=10^{18.70}$ (No.9). 
We also show the upper limits assuming a point source of $J_{\rm tot}=10^{19.15}$ with the shortest dash-dotted lines. 
 The center (right) column shows the achievable sensitivities assuming the same profiles of the 16 sources in Table~\ref{tabref} at 40kpc (160kpc) from the Earth. In each column, ${\it top}$, ${\it middle}$ and ${\it bottom}$ panel correspond to the upper limits on annihilation cross-sections of DM into $\bar{b}b$, $W^+W^-$ and $\tau^+\tau^-$, respectively.} 
\label{fig:sveach}
\end{center}
\end{figure*}
We conduct likelihood analyses on the simulated 500-hour observation of a dSph with DM density profiles listed in Table~\ref{tabref}. ${\it Top}$, ${\it middle}$ and ${\it bottom}$ rows in Figure~\ref{fig:sveach} show the cases of DM annihilating into $\bar{b}b$, $W^+W^-$ and $\tau^+\tau^-$, respectively. Panels in the left column show the sensitivities assuming the DM density profiles, distances, and J-factors ($J_{\rm tot}$) in Table~\ref{tabref}. Each line is the 95\% level upper limit corresponding to the profile in Figure~\ref{fig:rdependence}. Upper limits assuming profile No.14 (NFW model 5 in Ref.~\cite{Mashchenko:2005bj}) are the strongest while No.16 (PL of index 0 + cutoff model in Ref.~\cite{SanchezConde:2007te}) are the weakest in our sample. Other profiles give the upper limits in the shaded regions of the Figure~\ref{fig:sveach}, like middle-dash dotted lines corresponding to the cases of No.9. Sensitivities with a point source of $\log_{10}(J)=19.15$, which is the same as the J-factor of profile No. 14,  are also shown in a thin dash-dotted line. If we assume a point source, the upper limit always gets stronger. With the angular resolution of CTA, extended source structures are clearly resolved. Our results are consistent with the analytical discussion in Ref.~\cite{Ambrogi:2018skq}. Comparing between annihilation channels, wider regions of the cross-section parameter space can be covered for DM annihilating into $\tau^+\tau^-$ than for $\bar{b}b$ or $W^+W^-$ channels. This is due to the hard spectral feature which can be seen in the right panel of the Figure~\ref{srcspectrum}. The tendency is consistent with the latest results in Ref.~\cite{Abdalla:2018mve}, who assume line+broad spectra in specific WIMP models. Features in the sensitivity curves in Figure~\ref{fig:sveach} at $m_{\rm DM}={\cal O}(10)-{\cal O}(100)$ TeV result from the properties of the telescope. Center (Right) columns show the sensitivities for sources at smaller (larger) distances. We adopt the same distance among the profiles here. Differences between profiles are larger (smaller) for cases assuming 40 (160) kpc due to the angular extensions. In each panel, we also show the current limit by {\it Fermi} using 25 dSphs~\cite{Ackermann:2015zua} with a solid line and the expected sensitivities of the Galactic halo observations using CTA with dashed lines~\cite{Acharya:2017ttl,Pierre:2014tra,Wood:2013taa,Silverwood:2014yza,Roszkowski:2014iqa}. We show two cases assuming different DM density profiles for the expectations of G.C. observations because the DM density profile there is under discussion. The accessible annihilation cross-section is about two orders-of-magnitude smaller for the case assuming the Einasto profile (short-dashed line) than that assuming the Burkert profile (long-dashed line) as shown in these figures. 

We can expect better constraints when we adopt profiles based on the latest and detailed modelings of the dSphs. For example, we do not include the contribution from subhalos in dSphs since it is still under discussion (e.g.~\cite{Hutten:2016jko,Sanchez-Conde:2013yxa,Moline:2016pbm,Stref:2016uzb,Hiroshima:2018kfv,Charbonnier:2012gf,Bonnivard:2015pia,Hutten:2018aix}). Subhalos should enhance the annihilation signal, although little subhalo boost is expected in dSphs. Still, our results in this work provide conservative estimates.
\begin{figure*}[th!]
\begin{center}
\begin{tabular}{c} 
\begin{minipage}{0.33\hsize}
\begin{center}
\includegraphics[width=6cm]{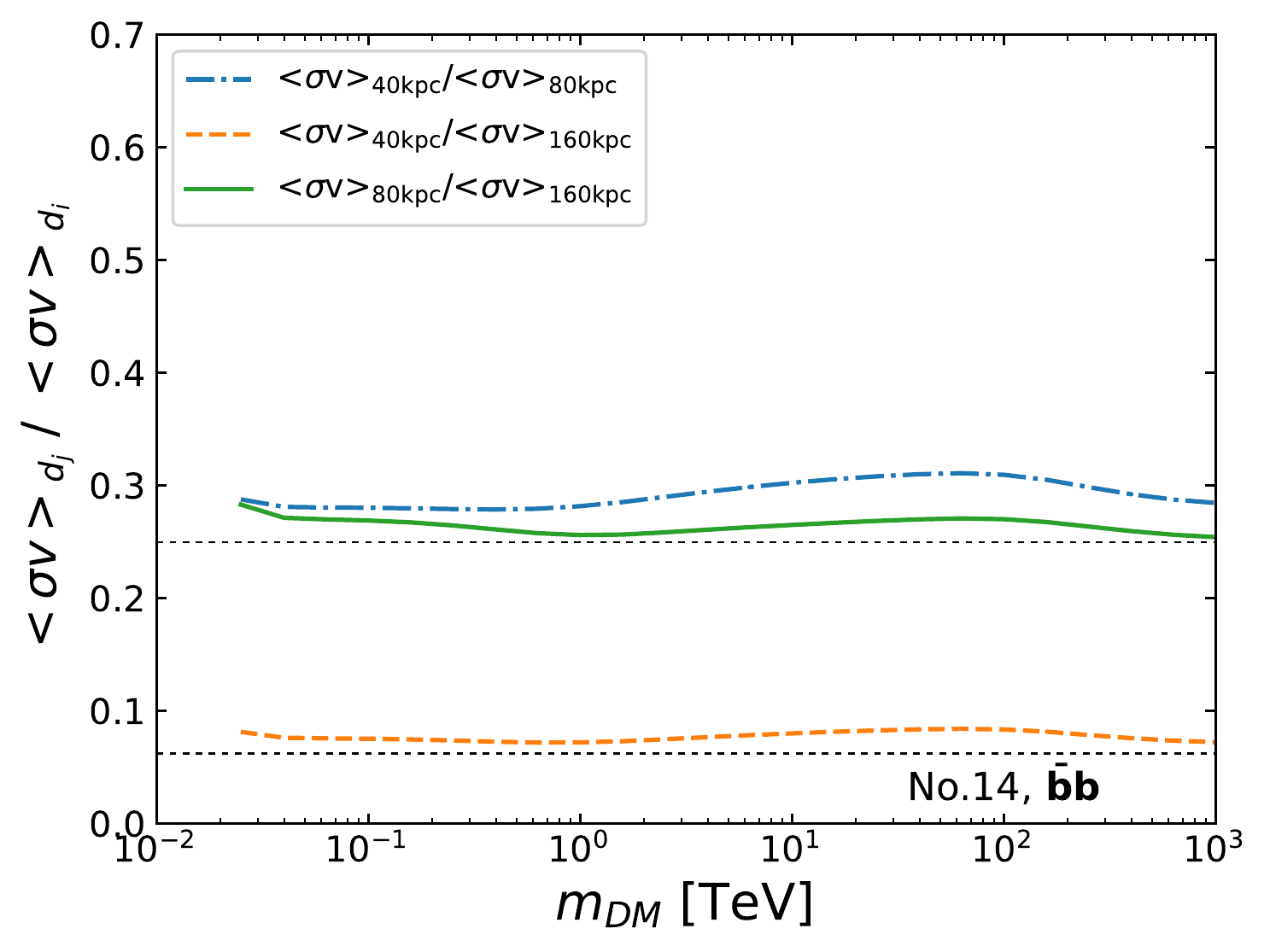} \hspace{0.8cm}
\end{center}
\end{minipage}
\begin{minipage}{0.33\hsize}
\begin{center}
\includegraphics[width=6cm]{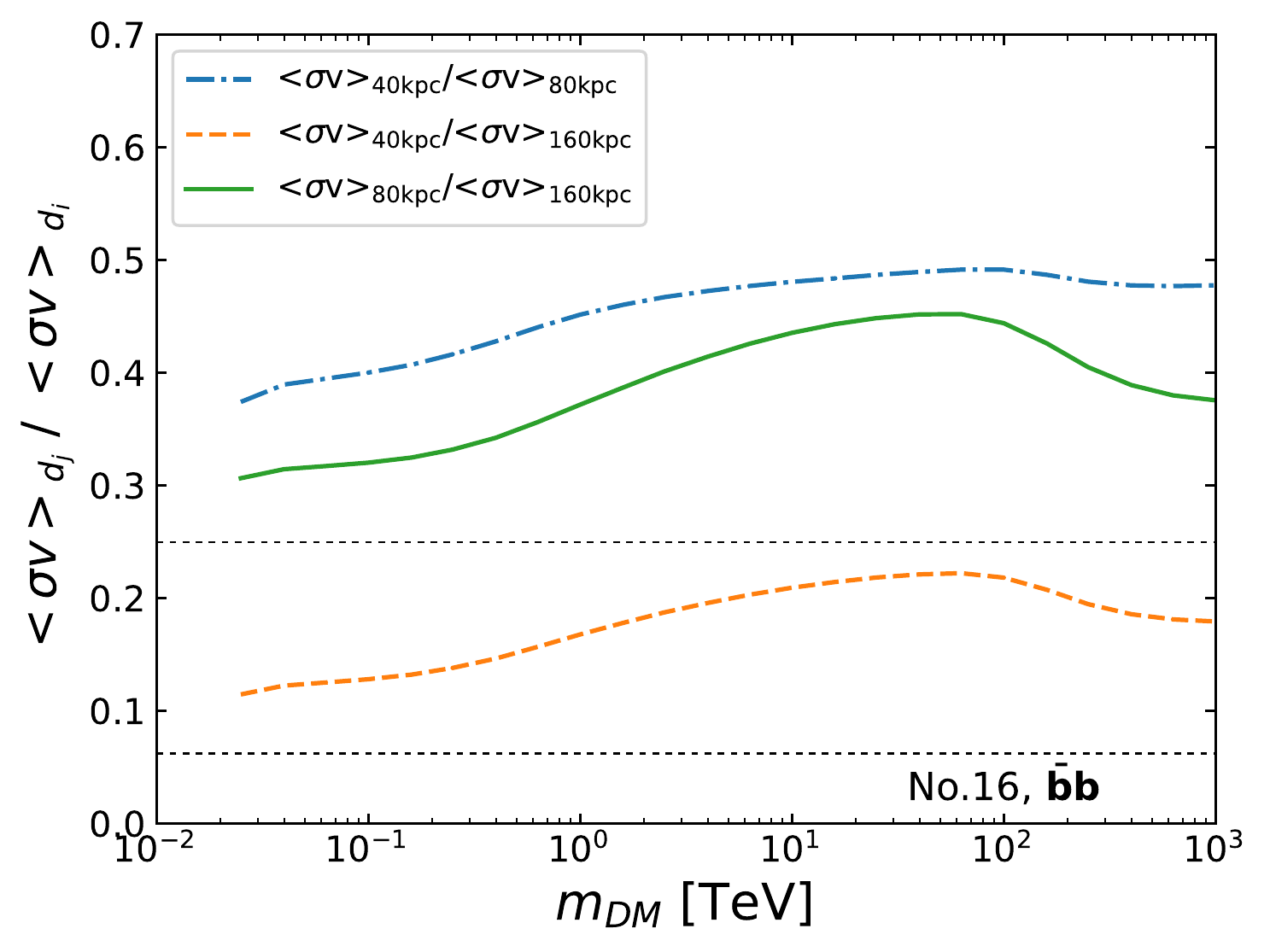} \hspace{0.8cm}
\end{center}
\end{minipage} \\
\begin{minipage}{0.33\hsize}
\begin{center}
\includegraphics[width=6cm]{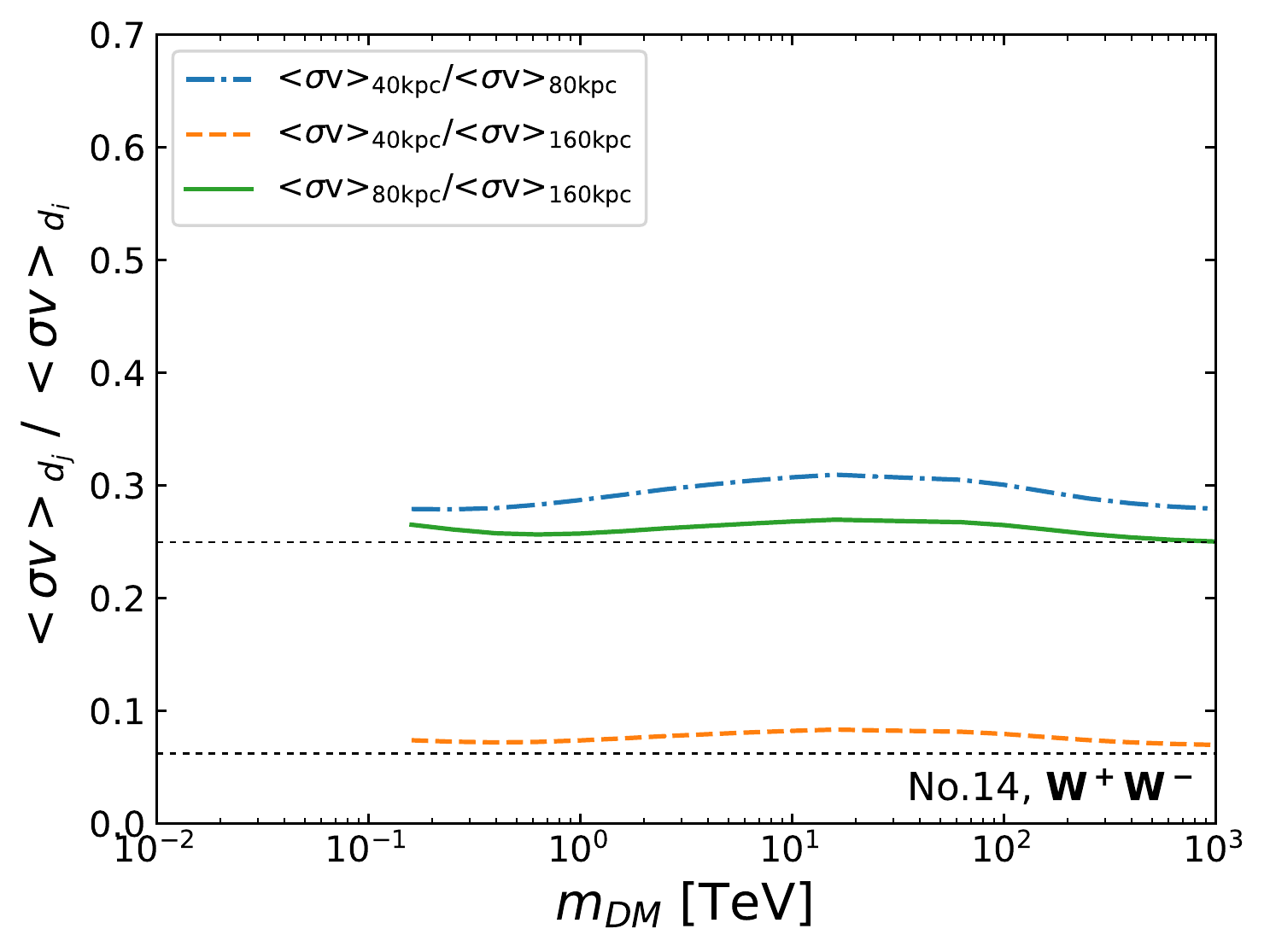} \hspace{0.8cm}
\end{center}
\end{minipage} 
\begin{minipage}{0.33\hsize}
\begin{center}
\includegraphics[width=6cm]{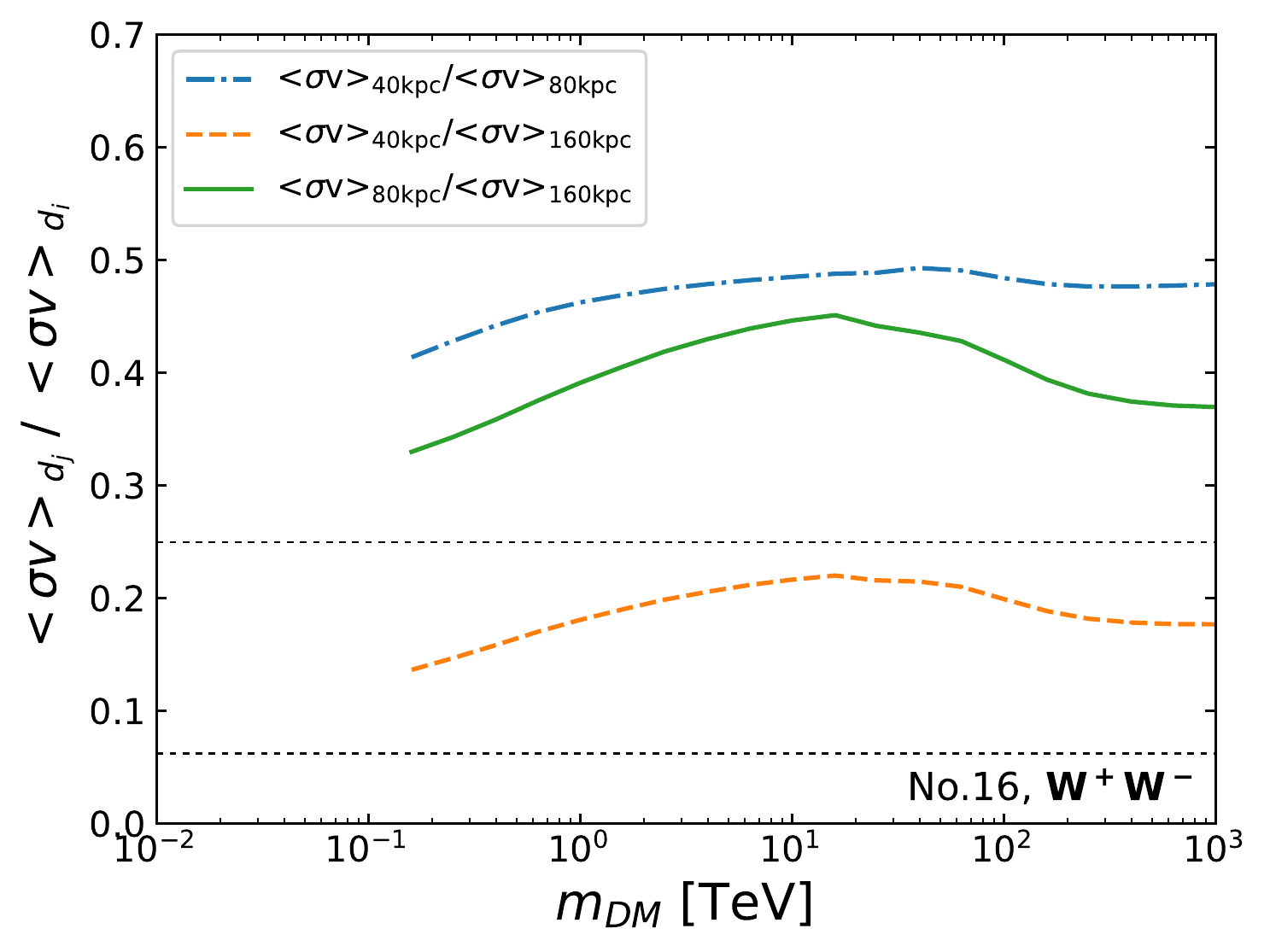} \hspace{0.8cm}
\end{center}
\end{minipage} \\
\begin{minipage}{0.33\hsize}
\begin{center}
\includegraphics[width=6cm]{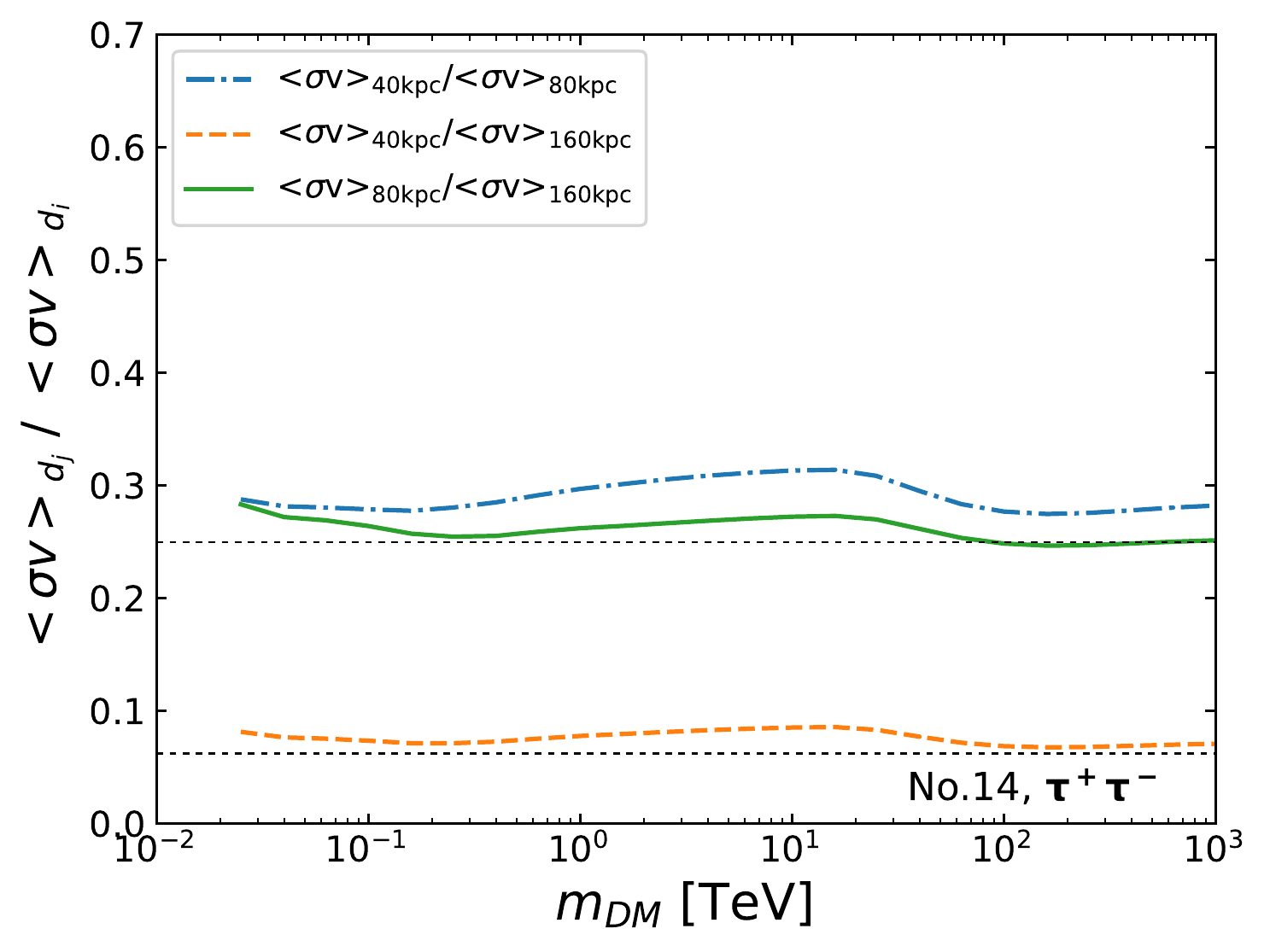} \hspace{0.8cm}
\end{center}
\end{minipage}
\begin{minipage}{0.33\hsize}
\begin{center}
\includegraphics[width=6cm]{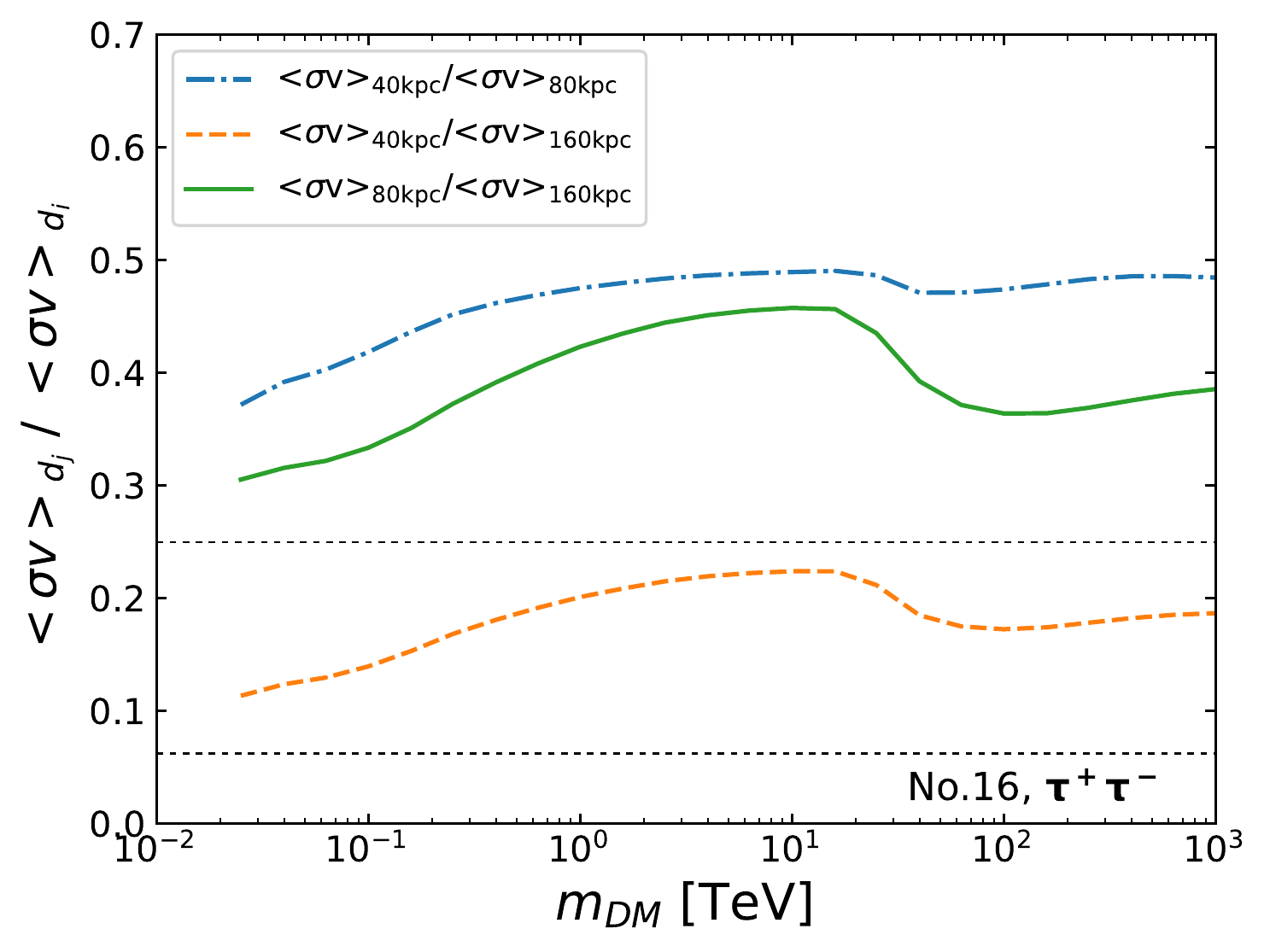} \hspace{0.8cm}
\end{center}
\end{minipage}
\end{tabular}
\caption{The ratio between the upper limits on the annihilation cross-section assuming the same profiles of dSphs at different distances, considering Profile No.14 (${\it left}$) and No.16 (${\it right}$). The expected value of ratio $\left<\sigma v\right>_{d_j}/\left<\sigma v\right>_{d_i}$ from the difference of J-factors are shown as dotted lines. The ratio of $(1/2)^2$ is expected for ($d_j$, $d_i$)=(40kpc, 80kpc) and (80kpc, 160kpc), while $(1/4)^2$ is expected for ($d_j$, $d_i$)=(40kpc, 160kpc). Deviations from the scaling from the J-factor are higher for the case of profile No.16.}
\label{fig:ratiodistvar}
\end{center}
\end{figure*}

In each channel of DM annihilation, the sensitivity achieves its best at $m_\mathrm{DM}=$ 630GeV, 1TeV and 250GeV for $\bar{b}b$, $W^+W^-$ and $\tau^+\tau^-$, respectively. These masses are universal among the profiles. 
By defining the rank of the profiles with the best points of the sensitivity in the DM mass range, we examine the relation between the annihilation channel final-state spectrum and the profile. There is no change in ranks of profiles between channels. No.14 in Table~\ref{tabref} is the strongest, No.16 is the weakest, and all other profiles lie between them in the same order. 

\begin{table*}[th!!]
\caption{J-factors ($J_\mathrm{tot}$) of profiles No.14 and No.16 in Table~\ref{tabref} assuming distance from the Earth to be 40kpc, 80kpc and 160kpc}
\begin{tabular}{cccc}\hline
Profile& \ \ log$_{10}$$J_\mathrm{tot}$ (40kpc) \ \ & \ \ log$_{10}$$J_\mathrm{tot}$ (80kpc) \ \ & \ \ log$_{10}$$J_\mathrm{tot}$ (160kpc) \ \ \\ \hline
No.14&19.79&19.17&18.53\\
No.16&19.18&18.58&17.98\\ \hline 
\end{tabular}
\label{tab:distvar}
\end{table*}

The dependence on the distance to the source is clarified in Figure~\ref{fig:ratiodistvar}. Assuming the source of profile No.14 and No.16 at 80kpc, 40kpc, 160kpc, we calculate the sensitivity and take the ratio of the upper limits on the annihilation cross-section. A source distance of 80kpc is chosen to be consistent with the distance for each model within the 1-$\sigma$ error. Corresponding J-factors are shown in Table~\ref{tab:distvar}, which show good agreements with the scaling law of $J\propto d^{-2}$ for point sources~\cite{Pace:2018tin,Evans:2016xwx}. The left (right) column corresponds to the profile No.14 (No.16). Profile No.14 (${\it left}$) almost follows the ratio expected from the scaling of J-factors in Table~\ref{tab:distvar}, while profile No.16 (${\it right}$) does not. For profile No.16, upper limits on $\left<\sigma v\right>$ get lower in milder ways than those expected from the scaling of the J-factor. Also, the DM mass dependence of the ratio differs between annihilation channels. 

\section{Discussion}
\label{sec:discussion}
\subsection{Dependences on profiles}
\label{ssec:profiledepends}
The difference of $\left<\sigma v\right>_{\mathrm{UL}}$ between profiles (short-dashed and long-dashed lines in Figure~\ref{fig:sveach}, for example) is caused by two effects. Subscript ``UL" denotes the upper limit here.
The values of $J_{\rm tot}$ affect the sensitivity to the annihilation cross-section $\left<\sigma v\right>_{\rm UL}$ in a direct way like cases analysing point sources with different J-factors. For analyses of extended sources, upper limits on the $\gamma$-ray flux $\phi_{\rm UL}$ are also affected by the details of DM density profiles hence $\left<\sigma v\right>_{\rm UL}$ is determined by combinations of these effects. The width of the shaded regions in Figure~\ref{fig:sveach} corresponds to this fact. When sources are at large distances (e.g. $d=$160kpc compared to $d=80$kpc), their density profile could not be resolved. Then the behaviour of the sensitivity curve becomes like that of a point source.  

\begin{figure}
\begin{center}
\includegraphics[width=8.5cm]{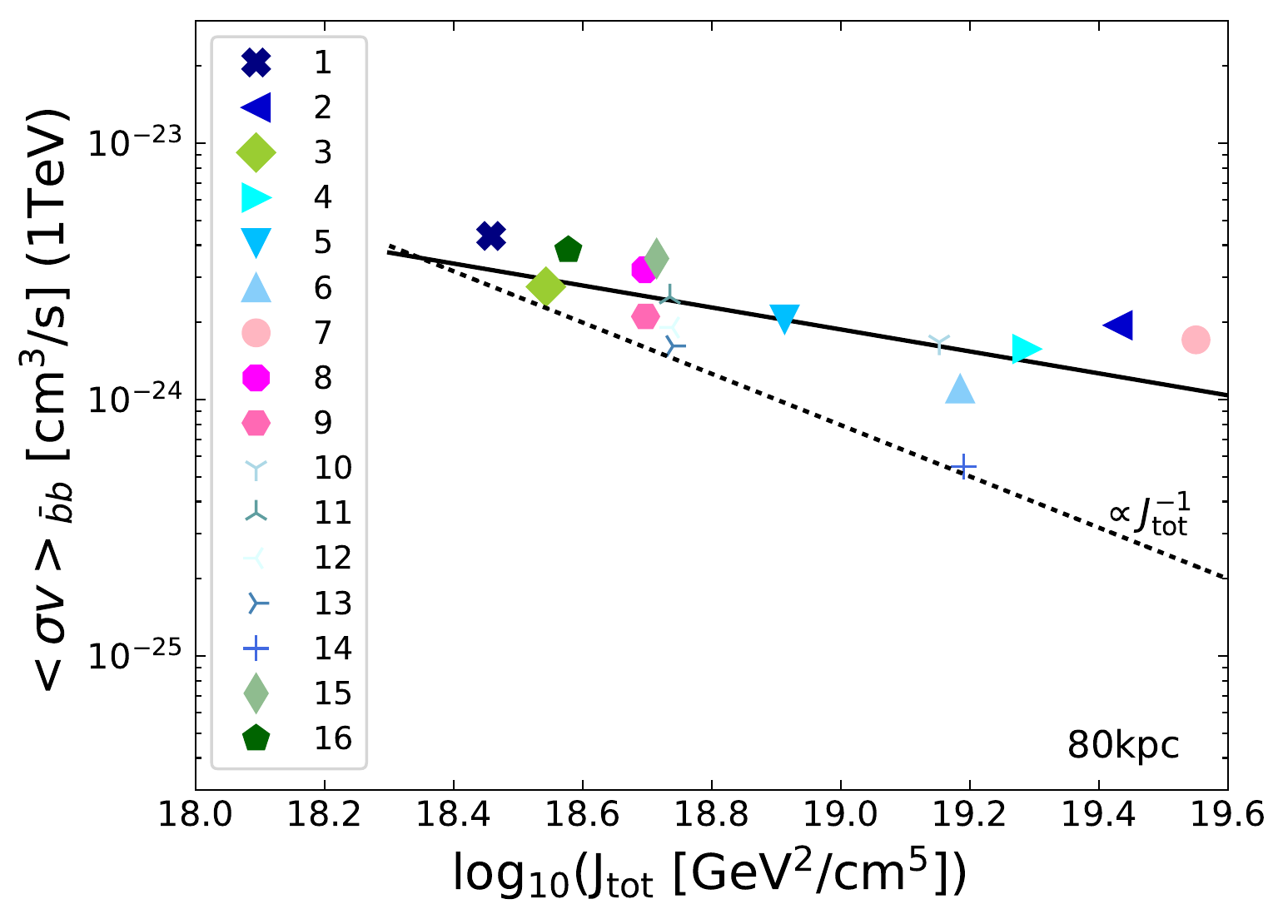} \hspace{0.8cm}
\caption{Upper limits on the cross-section for annihilation into $\bar{b}b$ pairs shown as a function of $J_{\rm tot}$. $\left<\sigma v\right>_{\bar{b}b}$ is evaluated at $m_\mathrm{DM}=630$GeV. The dotted line corresponds to the relation $\left<\sigma v\right>_{\rm UL}$ $\propto$ $J^{-1}$ which is expected for point sources. }
\label{fig:Jvssv20}
\end{center}
\end{figure}

To clarify this point, we show the relation between $\left<\sigma v\right>_{\rm UL}$ and $J_{\rm tot}$ in Figure~\ref{fig:Jvssv20}. $\left<\sigma v\right>_{\rm UL}$ is evaluated with $\left<\sigma v\right>_{\bar{b}b}$ at $m_{\rm DM}$=630GeV. Each marker corresponds to a profile in Table ~\ref{tabref}. The relation derived for cases of $W^+W^-$ or $\tau^+\tau^-$ is similar. The obtained $\left<\sigma v\right>_{\rm UL}$ does not follow the inverse of $J_{\rm tot}$, which is different from the case of $\phi_{\rm UL}$ independent of the DM distribution in dSphs like analyses of point sources. Therefore, a better understanding of the DM density profile is required in determining the goodness of the targets.
 We also investigate the dependence of the resultant limits on the DM density profile parameters. We search the relation between the upper limits of the annihilation cross-section $\left<\sigma v\right>$ and the DM density at a certain radius (0.1$^\circ$, 0.3$^\circ$, 0.5$^\circ$ and 1.0$^\circ$), the scale radius $r_s$, or the index $\gamma$ defined as $\rho(r)\propto r^{-\gamma}$ at the inner part ($r<r_s$). We find no correlations between either of the parameters and the achievable upper limits. Hence none of the single profile parameters can be used to select the target dSphs and we should select the targets based on the whole properties of their profiles.
\begin{figure}
\begin{center}
\includegraphics[width=8cm]{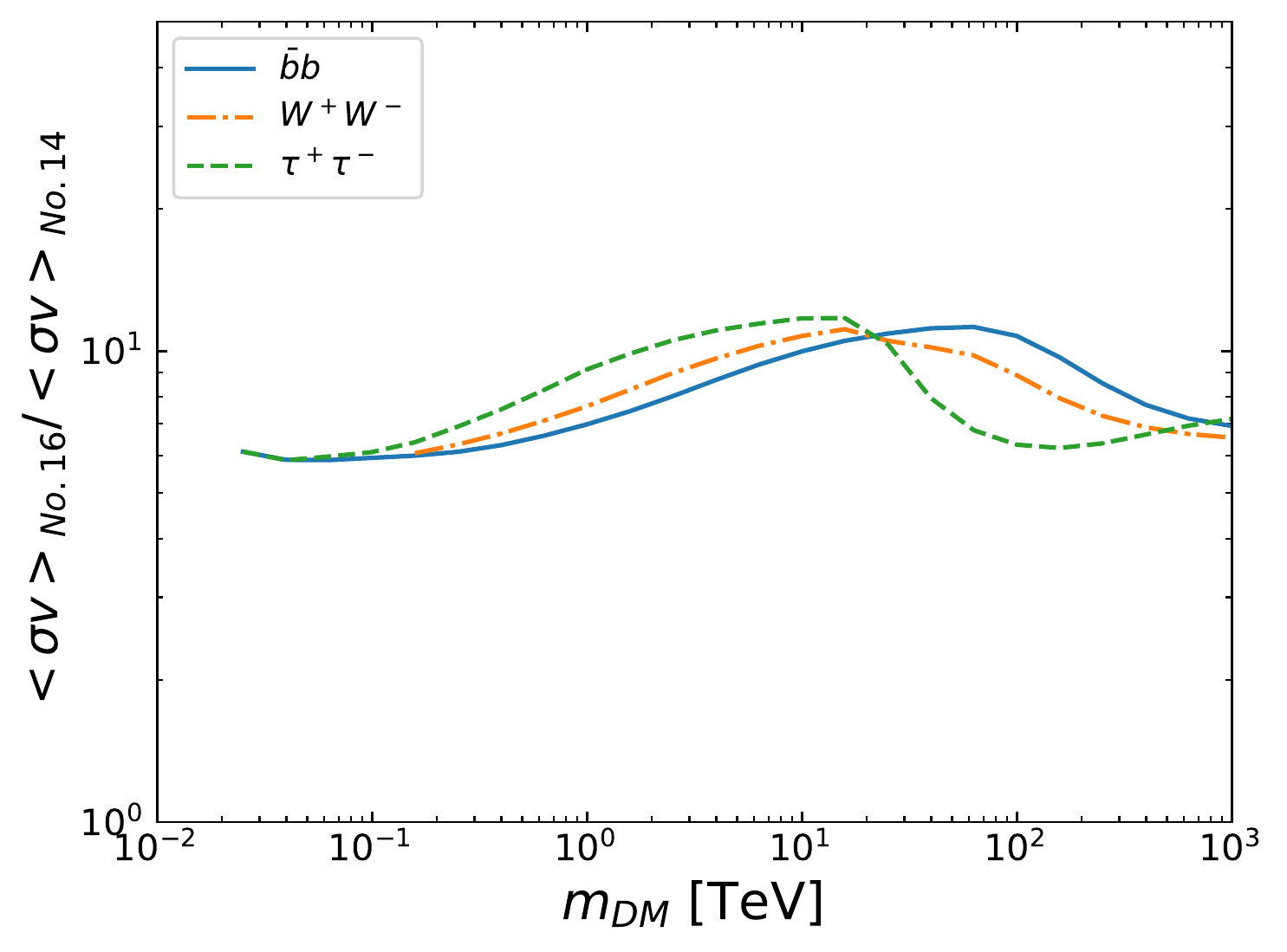} \hspace{0.8cm}
\caption{Ratios of upper limits on the annihilation cross-section obtained with profile No.16 and No.14 in Table~\ref{tabref}. The horizontal axis is the WIMP mass $m_\mathrm{DM}$ and the vertical axis is the ratio $r=\left<\sigma v\right>_\mathrm{No.16}/\left<\sigma v\right>_\mathrm{No.14}$. Solid, dot-dashed and dashed lines correspond  to the case considering the DM annihilating into $\bar{b}b$, $W^+W^-$ and $\tau^+\tau^-$, respectively.}
\label{fig:svratio}
\end{center}
\end{figure}

The dependence on the DM density profile also appears in the shape of the sensitivity curve. Figure~\ref{fig:svratio} shows the ratio of the obtained upper limits assuming profile No.16 and No.14 in Table~\ref{tabref}, $\left<\sigma v\right>_{No.16}/\left<\sigma v\right>_{No.14}$. For each annihilation channel, the ratio is about $\sim$10 and depending on the DM mass. A broad bump of the ratio at around $m_{\rm DM}$$\sim10$TeV to $\sim$500TeV is seen in the case of ${\bar{b}b}$, while a dip at around $m_{\rm DM}\sim$20TeV appears in the ratio for $\tau^+\tau^-$. For $W^+W^-$, a broad bump ranges from ${\cal O}(1)$ TeV to a few hundreds of TeV and it peaks at $m_{\rm DM}\lesssim{\cal O}(100)$TeV. The DM masses at around the bumps correspond to the $\gamma$-ray emission peaks of $E_\gamma\sim1-10$TeV (see Figure~\ref{srcspectrum}). The presence of the bumps can be interpreted as follows. The angular resolution of the CTA facility gets better as the energy increases. For example, it corresponds to about 0.1 degrees at $E_\gamma\sim$ 200 GeV and improves to about 0.04 degrees at $\gtrsim$1 TeV~\cite{Hassan:2017paq}. Therefore, the changes of $\phi_{\rm UL}$ are more significant at higher energies. On the other hand, in the very-high-energy regime at $E_\gamma>$10 TeV, almost no residual background events are expected. In such a case, the sensitivity is more determined by the detected number of signal events rather than the signal-to-noise ratio. It is a so-called ``signal-dominant case". In such cases, the angular resolution contributes less to the $\phi_{\rm UL}$, and $\phi_{\rm UL}$ is less affected by the spatial extension of the source. Then $\left<\sigma v\right>_{No.16}$ can get close to the expected values for those of point sources. The behaviour of the ratio $\left<\sigma v\right>_{No.16}/\left<\sigma v\right>_{No.14}$ is a manifestation of these effects since profile No.14 almost corresponds to a point source. Combining those two effects, the ratio between the profiles has the bump structures seen in Figure~\ref{fig:svratio}.

\subsection{Dependences on distances}
If the same profiles of dSphs are located at different distances $d_j$ and $d_i(>d_j)$, the J-factor increases by a factor of $(d_i/d_j)^2$ for closer ones. Then $\left<\sigma v\right>_{\rm UL}$ should be simply improved by $(d_i/d_j)^2$ when the target objects are point sources. However, improvements of $\left<\sigma v\right>_{\rm UL}$ are less than those expected from the scaling of J-factors as shown in Figure~\ref{fig:ratiodistvar}, due to the changes in $\phi_{\rm UL}$. Deviations from the scaling of J-factors are higher in the analyses assuming cored targets. 
Bumps are clearly seen in the right panels of Fig.~\ref{fig:ratiodistvar}. They peak at $m_{\rm DM}\lesssim$ ${\cal O}$(100)TeV for $\bar{b}b$ or $W^+W^-$, while at ${\cal O}$(10) TeV for $\tau^+\tau^-$. 
The features correspond to a peak at $E_\gamma\lesssim 10$TeV in the annihilation spectrum and a similar explanation of Sec.~\ref{ssec:profiledepends} holds. The sources at closer distances become more spatially extended such that  $\phi_{\rm UL}$ gets worse at higher energies.  As a result, the ratios $\left<\sigma v\right>_{d_j}/\left<\sigma v\right>_{d_i}$ for heavier WIMPs are more deviated from the expectations for point-sources. Contributions from the noise get lower at the very-high-energy regime, and consequently at $E_\gamma>$10 TeV $\phi_{\rm UL}$ are less affected by the spatial sizes of the source. Combining those two effects, the ratio between the upper limits on annihilation cross-section assuming the same profiles of dSphs at different distances have the bump structures as shown in Figure~\ref{fig:ratiodistvar}.

 Possibilities of the uncertainties in dSph analyses due to the modelings of isotropic background events are discussed in Ref.~\cite{Calore:2018sdx}. In our analyses, the normalization of the background is fitted simultaneously with the dark matter signals hence the additional uncertainties due to the modelings of the background would not appear. However, the background events become Poisson-like at $E_\gamma\gtrsim$~a few TeV where we expect signals from DM of $m_{\rm DM}\gtrsim{\cal O}(1)$ TeV. This might induce additional uncertainties of which contributions are small compared to those in DM spatial distributions in target dSphs. We quantify this point in future works.
\section{Conclusion}
\label{sec:conclusion}
Dependences of the accessible regions of the DM annihilation cross-section $\left<\sigma v\right>$ on the density profile of dSphs have been examined and quantified.
Since the DM density profile of each dSph is still actively debated, we have taken those of Draco dSph in the literature as examples. Based on the likelihood analyses on simulated 500-hour observations with CTA assuming the 16 profiles, we have shown that the achievable upper limits on DM annihilation cross-sections are highly dependent on the details of the spatial extensions of target dSphs. We have revealed that the probable region of the annihilation cross-section can differ by a factor of $\sim$10 if we change the profile models. The dependence is different from the case of a point source whose merit is fully described with a single J-factor value. To extract information about the nature of DM from $\gamma$-ray observations with CTA, we therefore conclude that it is crucial to better constrain the density profiles of the targets.

The dependence of upper limits on the distance to the target dSphs have been also considered. J-factors get higher for closer targets if profiles are the same. However, achievable upper limits are always worse than those expected from the scaling of J-factors due to the larger spatial extensions of sources. This effect is significant at around $\gamma$-ray energies around 10TeV. At around the same energy, the effect of the spatial extension of targets is also apparent in the comparison between the annihilation channels. Improved angular resolution and the signal-dominant situation in the higher $\gamma$-ray energy regions determine the behaviour of the sensitivity curve in combination.

\acknowledgments
We appreciate the careful reading and helpful suggestions for our earlier draft by Moritz H\"{u}tten and Vitor de Souza. Also, we thank Nicolas Produit and Subir Saker for comments. 
This research has made use of the CTA instrument response functions provided by the CTA Consortium and Observatory, see \url{http://www.cta-observatory.org/science/cta-performance/}(version prod3b-v1) for more details. This work is partly supported by the JSPS KAKENHI Grant Number 
 JP17H01131 (K.K.), and MEXT KAKENHI Grant Numbers
JP15H05889, JP18H04594 (K.K.), and 18K03665 (M.H.).  This paper has gone through internal review by the CTA consortium.

\bibliographystyle{utphys}
\bibliography{refHHK}

\providecommand{\href}[2]{#2}\begingroup\raggedright\begin{thebibliography}{100}

\bibitem{Bergstrom:2000pn}
L.~Bergstr{\"o}m, ``{Nonbaryonic dark matter: Observational evidence and
  detection methods},''
  \href{http://dx.doi.org/10.1088/0034-4885/63/5/2r3}{{\em Rept. Prog. Phys.}
  {\bfseries 63} (2000) 793},
\href{http://arxiv.org/abs/hep-ph/0002126}{{\ttfamily arXiv:hep-ph/0002126
  [hep-ph]}}.

\bibitem{Bertone:2004pz}
G.~Bertone, D.~Hooper, and J.~Silk, ``{Particle dark matter: Evidence,
  candidates and constraints},''
  \href{http://dx.doi.org/10.1016/j.physrep.2004.08.031}{{\em Phys. Rept.}
  {\bfseries 405} (2005) 279--390},
\href{http://arxiv.org/abs/hep-ph/0404175}{{\ttfamily arXiv:hep-ph/0404175
  [hep-ph]}}.

\bibitem{Gaskins:2016cha}
J.~M. Gaskins, ``{A review of indirect searches for particle dark matter},''
  \href{http://dx.doi.org/10.1080/00107514.2016.1175160}{{\em Contemp. Phys.}
  {\bfseries 57} no.~4, (2016) 496--525},
\href{http://arxiv.org/abs/1604.00014}{{\ttfamily arXiv:1604.00014
  [astro-ph.HE]}}.

\bibitem{Zwicky:1933gu}
F.~Zwicky, ``{Die Rotverschiebung von extragalaktischen Nebeln},''
  \href{http://dx.doi.org/10.1007/s10714-008-0707-4}{{\em Helv. Phys. Acta}
  {\bfseries 6} (1933) 110--127}.
[Gen. Rel. Grav.41,207(2009)].

\bibitem{Zwicky2009}
F.~Zwicky, ``Republication of: The redshift of extragalactic nebulae,''
  \href{http://dx.doi.org/10.1007/s10714-008-0707-4}{{\em General Relativity
  and Gravitation} {\bfseries 41} no.~1, (Jan, 2009) 207--224}.
  \url{https://doi.org/10.1007/s10714-008-0707-4}.

\bibitem{vanAlbada:1984js}
T.~S. van Albada, J.~N. Bahcall, K.~Begeman, and R.~Sancisi, ``{The
  Distribution of Dark Matter in the Spiral Galaxy {NGC}-3198},''
\href{http://dx.doi.org/10.1086/163375}{{\em Astrophys. J.} {\bfseries 295}
  (1985) 305--313}.

\bibitem{Salucci:2002jg}
P.~Salucci and A.~Borriello, ``{The intriguing distribution of dark matter in
  galaxies},'' {\em Lect. Notes Phys.} {\bfseries 616} (2003) 66--77,
\href{http://arxiv.org/abs/astro-ph/0203457}{{\ttfamily arXiv:astro-ph/0203457
  [astro-ph]}}.

\bibitem{Barrena:2002dp}
R.~Barrena, A.~Biviano, M.~Ramella, E.~E. Falco, and S.~Seitz, ``{The dynamical
  status of the cluster of galaxies 1e0657-56},''
  \href{http://dx.doi.org/10.1051/0004-6361:20020244}{{\em Astron. Astrophys.}
  {\bfseries 386} (2002) 816},
\href{http://arxiv.org/abs/astro-ph/0202323}{{\ttfamily arXiv:astro-ph/0202323
  [astro-ph]}}.

\bibitem{Clowe:2003tk}
D.~Clowe, A.~Gonzalez, and M.~Markevitch, ``{Weak lensing mass reconstruction
  of the interacting cluster 1E0657-558: Direct evidence for the existence of
  dark matter},'' \href{http://dx.doi.org/10.1086/381970}{{\em Astrophys. J.}
  {\bfseries 604} (2004) 596--603},
\href{http://arxiv.org/abs/astro-ph/0312273}{{\ttfamily arXiv:astro-ph/0312273
  [astro-ph]}}.

\bibitem{Peebles:1982ff}
P.~J.~E. Peebles, ``{Large scale background temperature and mass fluctuations
  due to scale invariant primeval perturbations},''
\href{http://dx.doi.org/10.1086/183911}{{\em Astrophys. J.} {\bfseries 263}
  (1982) L1--L5}.

\bibitem{Ade:2015xua}
{\bfseries Planck} Collaboration, P.~A.~R. Ade {\em et~al.}, ``{Planck 2015
  results. XIII. Cosmological parameters},''
  \href{http://dx.doi.org/10.1051/0004-6361/201525830}{{\em Astron. Astrophys.}
  {\bfseries 594} (2016) A13},
\href{http://arxiv.org/abs/1502.01589}{{\ttfamily arXiv:1502.01589
  [astro-ph.CO]}}.

\bibitem{Komatsu:2010fb}
{\bfseries WMAP} Collaboration, E.~Komatsu {\em et~al.}, ``{Seven-Year
  Wilkinson Microwave Anisotropy Probe (WMAP) Observations: Cosmological
  Interpretation},'' \href{http://dx.doi.org/10.1088/0067-0049/192/2/18}{{\em
  Astrophys. J. Suppl.} {\bfseries 192} (2011) 18},
\href{http://arxiv.org/abs/1001.4538}{{\ttfamily arXiv:1001.4538
  [astro-ph.CO]}}.

\bibitem{Akrami:2018vks}
{\bfseries Planck} Collaboration, Y.~Akrami {\em et~al.}, ``{Planck 2018
  results. I. Overview and the cosmological legacy of Planck},''
\href{http://arxiv.org/abs/1807.06205}{{\ttfamily arXiv:1807.06205
  [astro-ph.CO]}}.

\bibitem{Aghanim:2018eyx}
{\bfseries Planck} Collaboration, N.~Aghanim {\em et~al.}, ``{Planck 2018
  results. VI. Cosmological parameters},''
\href{http://arxiv.org/abs/1807.06209}{{\ttfamily arXiv:1807.06209
  [astro-ph.CO]}}.

\bibitem{Bringmann:2006mu}
T.~Bringmann and S.~Hofmann, ``{Thermal decoupling of WIMPs from first
  principles},'' \href{http://dx.doi.org/10.1088/1475-7516/2007/04/016,
  10.1088/1475-7516/2016/03/E02}{{\em JCAP} {\bfseries 0704} (2007) 016},
  \href{http://arxiv.org/abs/hep-ph/0612238}{{\ttfamily arXiv:hep-ph/0612238
  [hep-ph]}}.
[Erratum: JCAP1603,no.03,E02(2016)].

\bibitem{Rott:2012gh}
C.~Rott, ``{Review of Indirect WIMP Search Experiments},''
  \href{http://arxiv.org/abs/1210.4161}{{\ttfamily arXiv:1210.4161
  [astro-ph.HE]}}.
[Nucl. Phys. Proc. Suppl.235-236,413(2013)].

\bibitem{Mohapatra:1999gg}
R.~N. Mohapatra, F.~I. Olness, R.~Stroynowski, and V.~L. Teplitz, ``{Searching
  for strongly interacting massive particles (SIMPs)},''
  \href{http://dx.doi.org/10.1103/PhysRevD.60.115013}{{\em Phys. Rev.}
  {\bfseries D60} (1999) 115013},
\href{http://arxiv.org/abs/hep-ph/9906421}{{\ttfamily arXiv:hep-ph/9906421
  [hep-ph]}}.

\bibitem{Hochberg:2014dra}
Y.~Hochberg, E.~Kuflik, T.~Volansky, and J.~G. Wacker, ``{Mechanism for Thermal
  Relic Dark Matter of Strongly Interacting Massive Particles},''
  \href{http://dx.doi.org/10.1103/PhysRevLett.113.171301}{{\em Phys. Rev.
  Lett.} {\bfseries 113} (2014) 171301},
\href{http://arxiv.org/abs/1402.5143}{{\ttfamily arXiv:1402.5143 [hep-ph]}}.

\bibitem{Preskill:1982cy}
J.~Preskill, M.~B. Wise, and F.~Wilczek, ``{Cosmology of the Invisible
  Axion},''
\href{http://dx.doi.org/10.1016/0370-2693(83)90637-8}{{\em Phys. Lett.}
  {\bfseries 120B} (1983) 127--132}.

\bibitem{Rosenberg:2000wb}
L.~J. Rosenberg and K.~A. van Bibber, ``{Searches for invisible axions},''
\href{http://dx.doi.org/10.1016/S0370-1573(99)00045-9}{{\em Phys. Rept.}
  {\bfseries 325} (2000) 1--39}.

\bibitem{Visinelli:2009zm}
L.~Visinelli and P.~Gondolo, ``{Dark Matter Axions Revisited},''
  \href{http://dx.doi.org/10.1103/PhysRevD.80.035024}{{\em Phys. Rev.}
  {\bfseries D80} (2009) 035024},
\href{http://arxiv.org/abs/0903.4377}{{\ttfamily arXiv:0903.4377
  [astro-ph.CO]}}.

\bibitem{Dodelson:1993je}
S.~Dodelson and L.~M. Widrow, ``{Sterile-neutrinos as dark matter},''
  \href{http://dx.doi.org/10.1103/PhysRevLett.72.17}{{\em Phys. Rev. Lett.}
  {\bfseries 72} (1994) 17--20},
\href{http://arxiv.org/abs/hep-ph/9303287}{{\ttfamily arXiv:hep-ph/9303287
  [hep-ph]}}.

\bibitem{Shi:1998km}
X.-D. Shi and G.~M. Fuller, ``{A New dark matter candidate: Nonthermal sterile
  neutrinos},'' \href{http://dx.doi.org/10.1103/PhysRevLett.82.2832}{{\em Phys.
  Rev. Lett.} {\bfseries 82} (1999) 2832--2835},
\href{http://arxiv.org/abs/astro-ph/9810076}{{\ttfamily arXiv:astro-ph/9810076
  [astro-ph]}}.

\bibitem{Abazajian:2001nj}
K.~Abazajian, G.~M. Fuller, and M.~Patel, ``{Sterile neutrino hot, warm, and
  cold dark matter},'' \href{http://dx.doi.org/10.1103/PhysRevD.64.023501}{{\em
  Phys. Rev.} {\bfseries D64} (2001) 023501},
\href{http://arxiv.org/abs/astro-ph/0101524}{{\ttfamily arXiv:astro-ph/0101524
  [astro-ph]}}.

\bibitem{Boyarsky:2009ix}
A.~Boyarsky, O.~Ruchayskiy, and M.~Shaposhnikov, ``{The Role of sterile
  neutrinos in cosmology and astrophysics},''
  \href{http://dx.doi.org/10.1146/annurev.nucl.010909.083654}{{\em Ann. Rev.
  Nucl. Part. Sci.} {\bfseries 59} (2009) 191--214},
\href{http://arxiv.org/abs/0901.0011}{{\ttfamily arXiv:0901.0011 [hep-ph]}}.

\bibitem{Afshordi:2003zb}
N.~Afshordi, P.~McDonald, and D.~N. Spergel, ``{Primordial black holes as dark
  matter: The Power spectrum and evaporation of early structures},''
  \href{http://dx.doi.org/10.1086/378763}{{\em Astrophys. J.} {\bfseries 594}
  (2003) L71--L74},
\href{http://arxiv.org/abs/astro-ph/0302035}{{\ttfamily arXiv:astro-ph/0302035
  [astro-ph]}}.

\bibitem{Carr:2016drx}
B.~Carr, F.~Kuhnel, and M.~Sandstad, ``{Primordial Black Holes as Dark
  Matter},'' \href{http://dx.doi.org/10.1103/PhysRevD.94.083504}{{\em Phys.
  Rev.} {\bfseries D94} no.~8, (2016) 083504},
\href{http://arxiv.org/abs/1607.06077}{{\ttfamily arXiv:1607.06077
  [astro-ph.CO]}}.

\bibitem{Carr:2016hva}
B.~J. Carr, K.~Kohri, Y.~Sendouda, and J.~Yokoyama, ``{Constraints on
  primordial black holes from the Galactic gamma-ray background},''
  \href{http://dx.doi.org/10.1103/PhysRevD.94.044029}{{\em Phys. Rev.}
  {\bfseries D94} no.~4, (2016) 044029},
\href{http://arxiv.org/abs/1604.05349}{{\ttfamily arXiv:1604.05349
  [astro-ph.CO]}}.

\bibitem{Carr:2018rid}
B.~Carr and J.~Silk, ``{Primordial Black Holes as Seeds for Cosmic
  Structures},''
\href{http://arxiv.org/abs/1801.00672}{{\ttfamily arXiv:1801.00672
  [astro-ph.CO]}}.

\bibitem{Kohri:2018qtx}
K.~Kohri and T.~Terada, ``{Primordial Black Hole Dark Matter and LIGO/Virgo
  Merger Rate from Inflation with Running Spectral Indices},''
\href{http://arxiv.org/abs/1802.06785}{{\ttfamily arXiv:1802.06785
  [astro-ph.CO]}}.

\bibitem{Haber:1984rc}
H.~E. Haber and G.~L. Kane, ``{The Search for Supersymmetry: Probing Physics
  Beyond the Standard Model},''
\href{http://dx.doi.org/10.1016/0370-1573(85)90051-1}{{\em Phys. Rept.}
  {\bfseries 117} (1985) 75--263}.

\bibitem{Jungman:1995df}
G.~Jungman, M.~Kamionkowski, and K.~Griest, ``{Supersymmetric dark matter},''
  \href{http://dx.doi.org/10.1016/0370-1573(95)00058-5}{{\em Phys. Rept.}
  {\bfseries 267} (1996) 195--373},
\href{http://arxiv.org/abs/hep-ph/9506380}{{\ttfamily arXiv:hep-ph/9506380
  [hep-ph]}}.

\bibitem{Edsjo:1997bg}
J.~Edsjo and P.~Gondolo, ``{Neutralino relic density including
  coannihilations},'' \href{http://dx.doi.org/10.1103/PhysRevD.56.1879}{{\em
  Phys. Rev.} {\bfseries D56} (1997) 1879--1894},
\href{http://arxiv.org/abs/hep-ph/9704361}{{\ttfamily arXiv:hep-ph/9704361
  [hep-ph]}}.

\bibitem{Feng:2000gh}
J.~L. Feng, K.~T. Matchev, and F.~Wilczek, ``{Neutralino dark matter in focus
  point supersymmetry},''
  \href{http://dx.doi.org/10.1016/S0370-2693(00)00512-8}{{\em Phys. Lett.}
  {\bfseries B482} (2000) 388--399},
\href{http://arxiv.org/abs/hep-ph/0004043}{{\ttfamily arXiv:hep-ph/0004043
  [hep-ph]}}.

\bibitem{Giudice:2004tc}
G.~F. Giudice and A.~Romanino, ``{Split supersymmetry},''
  \href{http://dx.doi.org/10.1016/j.nuclphysb.2004.11.048,
  10.1016/j.nuclphysb.2004.08.001}{{\em Nucl. Phys.} {\bfseries B699} (2004)
  65--89}, \href{http://arxiv.org/abs/hep-ph/0406088}{{\ttfamily
  arXiv:hep-ph/0406088 [hep-ph]}}.
[Erratum: Nucl. Phys.B706,487(2005)].

\bibitem{Steigman:2012nb}
G.~Steigman, B.~Dasgupta, and J.~F. Beacom, ``{Precise Relic WIMP Abundance and
  its Impact on Searches for Dark Matter Annihilation},''
  \href{http://dx.doi.org/10.1103/PhysRevD.86.023506}{{\em Phys. Rev.}
  {\bfseries D86} (2012) 023506},
\href{http://arxiv.org/abs/1204.3622}{{\ttfamily arXiv:1204.3622 [hep-ph]}}.

\bibitem{Aaboud:2017aeu}
{\bfseries ATLAS} Collaboration, M.~Aaboud {\em et~al.}, ``{Search for
  top-squark pair production in final states with one lepton, jets, and missing
  transverse momentum using 36 fb$^{-1}$ of $\sqrt{s}=13$ TeV pp collision data
  with the ATLAS detector},''
\href{http://arxiv.org/abs/1711.11520}{{\ttfamily arXiv:1711.11520 [hep-ex]}}.

\bibitem{Sirunyan:2017xgm}
{\bfseries CMS} Collaboration, A.~M. Sirunyan {\em et~al.}, ``{Search for dark
  matter produced in association with heavy-flavor quark pairs in proton-proton
  collisions at $\sqrt{s}=13$ TeV},''
  \href{http://dx.doi.org/10.1140/epjc/s10052-017-5317-4}{{\em Eur. Phys. J.}
  {\bfseries C77} no.~12, (2017) 845},
\href{http://arxiv.org/abs/1706.02581}{{\ttfamily arXiv:1706.02581 [hep-ex]}}.

\bibitem{Akerib:2016vxi}
{\bfseries LUX} Collaboration, D.~S. Akerib {\em et~al.}, ``{Results from a
  search for dark matter in the complete LUX exposure},''
  \href{http://dx.doi.org/10.1103/PhysRevLett.118.021303}{{\em Phys. Rev.
  Lett.} {\bfseries 118} no.~2, (2017) 021303},
\href{http://arxiv.org/abs/1608.07648}{{\ttfamily arXiv:1608.07648
  [astro-ph.CO]}}.

\bibitem{Amole:2017dex}
{\bfseries PICO} Collaboration, C.~Amole {\em et~al.}, ``{Dark Matter Search
  Results from the PICO-60 C$_3$F$_8$ Bubble Chamber},''
  \href{http://dx.doi.org/10.1103/PhysRevLett.118.251301}{{\em Phys. Rev.
  Lett.} {\bfseries 118} no.~25, (2017) 251301},
\href{http://arxiv.org/abs/1702.07666}{{\ttfamily arXiv:1702.07666
  [astro-ph.CO]}}.

\bibitem{Aprile:2018dbl}
{\bfseries XENON} Collaboration, E.~Aprile {\em et~al.}, ``{Dark Matter Search
  Results from a One Tonne$\times$Year Exposure of XENON1T},''
\href{http://arxiv.org/abs/1805.12562}{{\ttfamily arXiv:1805.12562
  [astro-ph.CO]}}.

\bibitem{Fermi-LAT:2016uux}
{\bfseries DES, Fermi-LAT} Collaboration, A.~Albert {\em et~al.}, ``{Searching
  for Dark Matter Annihilation in Recently Discovered Milky Way Satellites with
  Fermi-LAT},'' \href{http://dx.doi.org/10.3847/1538-4357/834/2/110}{{\em
  Astrophys. J.} {\bfseries 834} no.~2, (2017) 110},
\href{http://arxiv.org/abs/1611.03184}{{\ttfamily arXiv:1611.03184
  [astro-ph.HE]}}.

\bibitem{Tremaine:1979we}
S.~Tremaine and J.~E. Gunn, ``{Dynamical Role of Light Neutral Leptons in
  Cosmology},'' \href{http://dx.doi.org/10.1103/PhysRevLett.42.407}{{\em Phys.
  Rev. Lett.} {\bfseries 42} (1979) 407--410}.
[,66(1979)].

\bibitem{Abazajian:2005xn}
K.~Abazajian, ``{Linear cosmological structure limits on warm dark matter},''
  \href{http://dx.doi.org/10.1103/PhysRevD.73.063513}{{\em Phys. Rev.}
  {\bfseries D73} (2006) 063513},
\href{http://arxiv.org/abs/astro-ph/0512631}{{\ttfamily arXiv:astro-ph/0512631
  [astro-ph]}}.

\bibitem{Horiuchi:2013noa}
S.~Horiuchi, P.~J. Humphrey, J.~Onorbe, K.~N. Abazajian, M.~Kaplinghat, and
  S.~Garrison-Kimmel, ``{Sterile neutrino dark matter bounds from galaxies of
  the Local Group},'' \href{http://dx.doi.org/10.1103/PhysRevD.89.025017}{{\em
  Phys. Rev.} {\bfseries D89} no.~2, (2014) 025017},
\href{http://arxiv.org/abs/1311.0282}{{\ttfamily arXiv:1311.0282
  [astro-ph.CO]}}.

\bibitem{Boyarsky:2008xj}
A.~Boyarsky, J.~Lesgourgues, O.~Ruchayskiy, and M.~Viel, ``{Lyman-alpha
  constraints on warm and on warm-plus-cold dark matter models},''
  \href{http://dx.doi.org/10.1088/1475-7516/2009/05/012}{{\em JCAP} {\bfseries
  0905} (2009) 012},
\href{http://arxiv.org/abs/0812.0010}{{\ttfamily arXiv:0812.0010 [astro-ph]}}.

\bibitem{Verzi:2017hro}
V.~Verzi, D.~Ivanov, and Y.~Tsunesada, ``{Measurement of Energy Spectrum of
  Ultra-High Energy Cosmic Rays},''
  \href{http://dx.doi.org/10.1093/ptep/ptx082}{{\em PTEP} {\bfseries 2017}
  no.~12, (2017) 12A103},
\href{http://arxiv.org/abs/1705.09111}{{\ttfamily arXiv:1705.09111
  [astro-ph.HE]}}.

\bibitem{Aab:2018chp}
{\bfseries Pierre Auger} Collaboration, A.~Aab {\em et~al.}, ``{An Indication
  of anisotropy in arrival directions of ultra-high-energy cosmic rays through
  comparison to the flux pattern of extragalactic gamma-ray sources},''
  \href{http://dx.doi.org/10.3847/2041-8213/aaa66d}{{\em Astrophys. J.}
  {\bfseries 853} no.~2, (2018) L29},
\href{http://arxiv.org/abs/1801.06160}{{\ttfamily arXiv:1801.06160
  [astro-ph.HE]}}.

\bibitem{Aartsen:2017sml}
{\bfseries IceCube} Collaboration, M.~G. Aartsen {\em et~al.}, ``{Neutrinos and
  Cosmic Rays Observed by IceCube},''
\href{http://arxiv.org/abs/1701.03731}{{\ttfamily arXiv:1701.03731
  [astro-ph.HE]}}.

\bibitem{Abeysekara:2017hyn}
A.~U. Abeysekara {\em et~al.}, ``{The 2HWC HAWC Observatory Gamma Ray
  Catalog},'' \href{http://dx.doi.org/10.3847/1538-4357/aa7556}{{\em Astrophys.
  J.} {\bfseries 843} no.~1, (2017) 40},
\href{http://arxiv.org/abs/1702.02992}{{\ttfamily arXiv:1702.02992
  [astro-ph.HE]}}.

\bibitem{Acharya:2017ttl}
{\bfseries CTA Consortium} Collaboration, B.~S. Acharya {\em et~al.},
  \href{http://dx.doi.org/10.1142/10986}{{\em {Science with the Cherenkov
  Telescope Array}}}.
\newblock {WORLD SCIENTIFIC}, 2019.
\newblock \href{http://arxiv.org/abs/1709.07997}{{\ttfamily arXiv:1709.07997
  [astro-ph.IM]}}.
\newblock
\url{{https://www.worldscientific.com/doi/abs/10.1142/10986}}.
\newblock

\bibitem{Aguilar:2015ooa}
{\bfseries AMS} Collaboration, M.~Aguilar {\em et~al.}, ``{Precision
  Measurement of the Proton Flux in Primary Cosmic Rays from Rigidity 1 GV to
  1.8 TV with the Alpha Magnetic Spectrometer on the International Space
  Station},''
\href{http://dx.doi.org/10.1103/PhysRevLett.114.171103}{{\em Phys. Rev. Lett.}
  {\bfseries 114} (2015) 171103}.

\bibitem{TheDAMPE:2017dtc}
{\bfseries DAMPE} Collaboration, J.~Chang {\em et~al.}, ``{The DArk Matter
  Particle Explorer mission},''
  \href{http://dx.doi.org/10.1016/j.astropartphys.2017.08.005}{{\em Astropart.
  Phys.} {\bfseries 95} (2017) 6--24},
\href{http://arxiv.org/abs/1706.08453}{{\ttfamily arXiv:1706.08453
  [astro-ph.IM]}}.

\bibitem{Motz:2015cua}
H.~Motz, Y.~Asaoka, S.~Torii, and S.~Bhattacharyya, ``{CALET's Sensitivity to
  Dark Matter Annihilation in the Galactic Halo},''
  \href{http://dx.doi.org/10.1088/1475-7516/2015/12/047}{{\em JCAP} {\bfseries
  1512} no.~12, (2015) 047},
\href{http://arxiv.org/abs/1510.03168}{{\ttfamily arXiv:1510.03168
  [astro-ph.HE]}}.

\bibitem{Aartsen:2017ulx}
{\bfseries IceCube} Collaboration, M.~G. Aartsen {\em et~al.}, ``{Search for
  Neutrinos from Dark Matter Self-Annihilations in the center of the Milky Way
  with 3 years of IceCube/DeepCore},''
  \href{http://dx.doi.org/10.1140/epjc/s10052-017-5213-y}{{\em Eur. Phys. J.}
  {\bfseries C77} no.~9, (2017) 627},
\href{http://arxiv.org/abs/1705.08103}{{\ttfamily arXiv:1705.08103 [hep-ex]}}.

\bibitem{Albert:2016dsy}
{\bfseries ANTARES} Collaboration, A.~Albert {\em et~al.}, ``{Search for Dark
  Matter Annihilation in the Earth using the ANTARES Neutrino Telescope},''
  \href{http://dx.doi.org/10.1016/j.dark.2017.04.005}{{\em Phys. Dark Univ.}
  {\bfseries 16} (2017) 41--48},
\href{http://arxiv.org/abs/1612.06792}{{\ttfamily arXiv:1612.06792 [hep-ex]}}.

\bibitem{Posti2019}
L.~{Posti} and A.~{Helmi}, ``{Mass and shape of the Milky Way's dark matter
  halo with globular clusters from Gaia and Hubble},''
  \href{http://dx.doi.org/10.1051/0004-6361/201833355}{{\em Astron. Astrophys.}
  {\bfseries 621} (2019) A56},
  \href{http://arxiv.org/abs/astro-ph/0202323}{{\ttfamily
  arXiv:astro-ph/0202323 [astro-ph]}}.

\bibitem{Gondolo:1999ef}
P.~Gondolo and J.~Silk, ``{Dark matter annihilation at the galactic center},''
  \href{http://dx.doi.org/10.1103/PhysRevLett.83.1719}{{\em Phys. Rev. Lett.}
  {\bfseries 83} (1999) 1719--1722},
\href{http://arxiv.org/abs/astro-ph/9906391}{{\ttfamily arXiv:astro-ph/9906391
  [astro-ph]}}.

\bibitem{TheFermi-LAT:2017vmf}
{\bfseries Fermi-LAT} Collaboration, M.~Ackermann {\em et~al.}, ``{The Fermi
  Galactic Center GeV Excess and Implications for Dark Matter},''
  \href{http://dx.doi.org/10.3847/1538-4357/aa6cab}{{\em Astrophys. J.}
  {\bfseries 840} no.~1, (2017) 43},
\href{http://arxiv.org/abs/1704.03910}{{\ttfamily arXiv:1704.03910
  [astro-ph.HE]}}.

\bibitem{HESS:2015cda}
{\bfseries H.E.S.S.} Collaboration, A.~Abramowski {\em et~al.}, ``{Constraints
  on an Annihilation Signal from a Core of Constant Dark Matter Density around
  the Milky Way Center with H.E.S.S.},''
  \href{http://dx.doi.org/10.1103/PhysRevLett.114.081301}{{\em Phys. Rev.
  Lett.} {\bfseries 114} no.~8, (2015) 081301},
\href{http://arxiv.org/abs/1502.03244}{{\ttfamily arXiv:1502.03244
  [astro-ph.HE]}}.

\bibitem{Pierre:2014tra}
M.~Pierre, J.~M. Siegal-Gaskins, and P.~Scott, ``{Sensitivity of CTA to dark
  matter signals from the Galactic Center},''
  \href{http://dx.doi.org/10.1088/1475-7516/2014/10/E01,
  10.1088/1475-7516/2014/06/024}{{\em JCAP} {\bfseries 1406} (2014) 024},
  \href{http://arxiv.org/abs/1401.7330}{{\ttfamily arXiv:1401.7330
  [astro-ph.HE]}}.
[Erratum: JCAP1410,E01(2014)].

\bibitem{Gammaldi:2016uhg}
V.~Gammaldi, V.~Avila-Reese, O.~Valenzuela, and A.~X. Gonzales-Morales,
  ``{Analysis of the very inner Milky Way dark matter distribution and
  gamma-ray signals},''
  \href{http://dx.doi.org/10.1103/PhysRevD.94.121301}{{\em Phys. Rev.}
  {\bfseries D94} no.~12, (2016) 121301},
\href{http://arxiv.org/abs/1607.02012}{{\ttfamily arXiv:1607.02012
  [astro-ph.HE]}}.

\bibitem{Taylor:2015jaa}
C.~Taylor, M.~Boylan-Kolchin, P.~Torrey, M.~Vogelsberger, and L.~Hernquist,
  ``{The Mass Profile of the Milky Way to the Virial Radius from the Illustris
  Simulation},'' \href{http://dx.doi.org/10.1093/mnras/stw1522}{{\em Mon. Not.
  Roy. Astron. Soc.} {\bfseries 461} no.~4, (2016) 3483--3493},
\href{http://arxiv.org/abs/1510.06409}{{\ttfamily arXiv:1510.06409
  [astro-ph.CO]}}.

\bibitem{Lake:1990du}
G.~Lake, ``{Detectability of gamma-rays from clumps of dark matter},''
\href{http://dx.doi.org/10.1038/346039a0}{{\em Nature} {\bfseries 346} (1990)
  39--40}.

\bibitem{Evans:2003sc}
N.~W. Evans, F.~Ferrer, and S.~Sarkar, ``{A 'Baedecker' for the dark matter
  annihilation signal},''
  \href{http://dx.doi.org/10.1103/PhysRevD.69.123501}{{\em Phys. Rev.}
  {\bfseries D69} (2004) 123501},
\href{http://arxiv.org/abs/astro-ph/0311145}{{\ttfamily arXiv:astro-ph/0311145
  [astro-ph]}}.

\bibitem{Martin:2015xla}
N.~F. Martin {\em et~al.}, ``{Hydra II: a faint and compact Milky Way dwarf
  galaxy found in the Survey of the Magellanic Stellar History},''
  \href{http://dx.doi.org/10.1088/2041-8205/804/1/L5}{{\em Astrophys. J.}
  {\bfseries 804} no.~1, (2015) L5},
\href{http://arxiv.org/abs/1503.06216}{{\ttfamily arXiv:1503.06216
  [astro-ph.GA]}}.

\bibitem{Kim:2015ila}
D.~Kim and H.~Jerjen, ``{Horologium II: a Second Ultra-faint Milky Way
  Satellite in the Horologium Constellation},''
  \href{http://dx.doi.org/10.1088/2041-8205/808/2/L39}{{\em Astrophys. J.}
  {\bfseries 808} (2015) L39},
\href{http://arxiv.org/abs/1505.04948}{{\ttfamily arXiv:1505.04948
  [astro-ph.GA]}}.

\bibitem{Laevens:2015una}
B.~P.~M. Laevens {\em et~al.}, ``{A New Faint Milky Way Satellite Discovered in
  the Pan-STARRS1 3 pi Survey},''
  \href{http://dx.doi.org/10.1088/2041-8205/802/2/L18}{{\em Astrophys. J.}
  {\bfseries 802} (2015) L18},
\href{http://arxiv.org/abs/1503.05554}{{\ttfamily arXiv:1503.05554
  [astro-ph.GA]}}.

\bibitem{Laevens:2015kla}
B.~P.~M. Laevens {\em et~al.}, ``{Sagittarius II, Draco II and Laevens 3: Three
  new Milky way Satellites Discovered in the Pan-starrs 1 3$\pi$ Survey},''
  \href{http://dx.doi.org/10.1088/0004-637X/813/1/44}{{\em Astrophys. J.}
  {\bfseries 813} no.~1, (2015) 44},
\href{http://arxiv.org/abs/1507.07564}{{\ttfamily arXiv:1507.07564
  [astro-ph.GA]}}.

\bibitem{Luque:2015txp}
{\bfseries DES} Collaboration, E.~Luque {\em et~al.}, ``{Digging Deeper into
  the Southern Skies: A Compact Milky-Way Companion Discovered in First-Year
  Dark Energy Survey Data},''
  \href{http://dx.doi.org/10.1093/mnras/stw302}{{\em Mon. Not. Roy. Astron.
  Soc.} {\bfseries 458} no.~1, (2016) 603--612},
\href{http://arxiv.org/abs/1508.02381}{{\ttfamily arXiv:1508.02381
  [astro-ph.GA]}}.

\bibitem{Koposov:2015jla}
S.~E. Koposov {\em et~al.}, ``{Kinematics and chemistry of recently discovered
  Reticulum 2 and Horologium 1 dwarf galaxies},''
  \href{http://dx.doi.org/10.1088/0004-637X/811/1/62}{{\em Astrophys. J.}
  {\bfseries 811} no.~1, (2015) 62},
\href{http://arxiv.org/abs/1504.07916}{{\ttfamily arXiv:1504.07916
  [astro-ph.GA]}}.

\bibitem{Bechtol:2015cbp}
{\bfseries DES} Collaboration, K.~Bechtol {\em et~al.}, ``{Eight New Milky Way
  Companions Discovered in First-Year Dark Energy Survey Data},''
  \href{http://dx.doi.org/10.1088/0004-637X/807/1/50}{{\em Astrophys. J.}
  {\bfseries 807} no.~1, (2015) 50},
\href{http://arxiv.org/abs/1503.02584}{{\ttfamily arXiv:1503.02584
  [astro-ph.GA]}}.

\bibitem{McConnachie:2012vd}
A.~W. McConnachie, ``{The observed properties of dwarf galaxies in and around
  the Local Group},'' \href{http://dx.doi.org/10.1088/0004-6256/144/1/4}{{\em
  Astron. J.} {\bfseries 144} (2012) 4},
\href{http://arxiv.org/abs/1204.1562}{{\ttfamily arXiv:1204.1562
  [astro-ph.CO]}}.

\bibitem{Gaia2018}
{\bfseries Gaia} Collaboration, ``{Gaia Data Release 2. Kinematics of globular
  clusters and dwarf galaxies around the Milky Way},''
  \href{http://dx.doi.org/10.1051/0004-6361/201832698}{{\em Astron. Astrophys.}
  {\bfseries 616} (2018) A12},
  \href{http://arxiv.org/abs/1804.09381}{{\ttfamily arXiv:1804.09381
  [astro-ph.GA]}}.

\bibitem{Simons2018}
J.~D. {Simon}, ``{Gaia Proper Motions and Orbits of the Ultra-faint Milky Way
  Satellites},'' \href{http://dx.doi.org/10.3847/1538-4357/aacdfb}{{\em
  Astrophys. J.} {\bfseries 863} (2018) 89},
  \href{http://arxiv.org/abs/1804.10230}{{\ttfamily arXiv:1804.10230
  [astro-ph.GA]}}.

\bibitem{Strigari:2006rd}
L.~E. Strigari, S.~M. Koushiappas, J.~S. Bullock, and M.~Kaplinghat, ``{Precise
  constraints on the dark matter content of Milky Way dwarf galaxies for
  gamma-ray experiments},''
  \href{http://dx.doi.org/10.1103/PhysRevD.75.083526}{{\em Phys. Rev.}
  {\bfseries D75} (2007) 083526},
\href{http://arxiv.org/abs/astro-ph/0611925}{{\ttfamily arXiv:astro-ph/0611925
  [astro-ph]}}.

\bibitem{Charbonnier:2011ft}
A.~Charbonnier {\em et~al.}, ``{Dark matter profiles and annihilation in dwarf
  spheroidal galaxies: prospectives for present and future gamma-ray
  observatories - I. The classical dSphs},''
  \href{http://dx.doi.org/10.1111/j.1365-2966.2011.19387.x}{{\em Mon. Not. Roy.
  Astron. Soc.} {\bfseries 418} (2011) 1526--1556},
\href{http://arxiv.org/abs/1104.0412}{{\ttfamily arXiv:1104.0412
  [astro-ph.HE]}}.

\bibitem{Hayashi:2016kcy}
K.~Hayashi, K.~Ichikawa, S.~Matsumoto, M.~Ibe, M.~N. Ishigaki, and H.~Sugai,
  ``{Dark matter annihilation and decay from non-spherical dark halos in
  galactic dwarf satellites},''
  \href{http://dx.doi.org/10.1093/mnras/stw1457}{{\em Mon. Not. Roy. Astron.
  Soc.} {\bfseries 461} no.~3, (2016) 2914--2928},
\href{http://arxiv.org/abs/1603.08046}{{\ttfamily arXiv:1603.08046
  [astro-ph.GA]}}.

\bibitem{Mateo:1998wg}
M.~Mateo, ``{Dwarf galaxies of the Local Group},''
  \href{http://dx.doi.org/10.1146/annurev.astro.36.1.435}{{\em Ann. Rev.
  Astron. Astrophys.} {\bfseries 36} (1998) 435--506},
\href{http://arxiv.org/abs/astro-ph/9810070}{{\ttfamily arXiv:astro-ph/9810070
  [astro-ph]}}.

\bibitem{Strigari:2007at}
L.~E. Strigari, S.~M. Koushiappas, J.~S. Bullock, M.~Kaplinghat, J.~D. Simon,
  M.~Geha, and B.~Willman, ``{The Most Dark Matter Dominated Galaxies:
  Predicted Gamma-ray Signals from the Faintest Milky Way Dwarfs},''
  \href{http://dx.doi.org/10.1086/529488}{{\em Astrophys. J.} {\bfseries 678}
  (2008) 614},
\href{http://arxiv.org/abs/0709.1510}{{\ttfamily arXiv:0709.1510 [astro-ph]}}.

\bibitem{Battaglia:2013wqa}
G.~Battaglia, A.~Helmi, and M.~Breddels, ``{Internal kinematics and dynamical
  models of dwarf spheroidal galaxies around the Milky Way},''
  \href{http://dx.doi.org/10.1016/j.newar.2013.05.003}{{\em New Astron. Rev.}
  {\bfseries 57} (2013) 52--79},
\href{http://arxiv.org/abs/1305.5965}{{\ttfamily arXiv:1305.5965
  [astro-ph.CO]}}.

\bibitem{Pace:2018tin}
A.~B. Pace and L.~E. Strigari, ``{Scaling Relations for Dark Matter
  Annihilation and Decay Profiles in Dwarf Spheroidal Galaxies},''
\href{http://arxiv.org/abs/1802.06811}{{\ttfamily arXiv:1802.06811
  [astro-ph.GA]}}.

\bibitem{Irwin:1995tb}
M.~Irwin and D.~Hatzidimitriou, ``{Structural parameters for the Galactic dwarf
  spheroidals},''
{\em Mon. Not. Roy. Astron. Soc.} {\bfseries 277} (1995) 1354--1378.

\bibitem{Geringer-Sameth:2015lua}
A.~Geringer-Sameth, M.~G. Walker, S.~M. Koushiappas, S.~E. Koposov,
  V.~Belokurov, G.~Torrealba, and N.~W. Evans, ``{Indication of Gamma-ray
  Emission from the Newly Discovered Dwarf Galaxy Reticulum II},''
  \href{http://dx.doi.org/10.1103/PhysRevLett.115.081101}{{\em Phys. Rev.
  Lett.} {\bfseries 115} no.~8, (2015) 081101},
\href{http://arxiv.org/abs/1503.02320}{{\ttfamily arXiv:1503.02320
  [astro-ph.HE]}}.

\bibitem{Abdo:2010ex}
{\bfseries Fermi-LAT} Collaboration, A.~A. Abdo {\em et~al.}, ``{Observations
  of Milky Way Dwarf Spheroidal galaxies with the Fermi-LAT detector and
  constraints on Dark Matter models},''
  \href{http://dx.doi.org/10.1088/0004-637X/712/1/147}{{\em Astrophys. J.}
  {\bfseries 712} (2010) 147--158},
\href{http://arxiv.org/abs/1001.4531}{{\ttfamily arXiv:1001.4531
  [astro-ph.CO]}}.

\bibitem{Ackermann:2011wa}
{\bfseries Fermi-LAT} Collaboration, M.~Ackermann {\em et~al.}, ``{Constraining
  Dark Matter Models from a Combined Analysis of Milky Way Satellites with the
  Fermi Large Area Telescope},''
  \href{http://dx.doi.org/10.1103/PhysRevLett.107.241302}{{\em Phys. Rev.
  Lett.} {\bfseries 107} (2011) 241302},
\href{http://arxiv.org/abs/1108.3546}{{\ttfamily arXiv:1108.3546
  [astro-ph.HE]}}.

\bibitem{Ackermann:2013yva}
{\bfseries Fermi-LAT} Collaboration, M.~Ackermann {\em et~al.}, ``{Dark matter
  constraints from observations of 25 Milky Way satellite galaxies with the
  Fermi Large Area Telescope},''
  \href{http://dx.doi.org/10.1103/PhysRevD.89.042001}{{\em Phys. Rev.}
  {\bfseries D89} (2014) 042001},
\href{http://arxiv.org/abs/1310.0828}{{\ttfamily arXiv:1310.0828
  [astro-ph.HE]}}.

\bibitem{Ackermann:2015zua}
{\bfseries Fermi-LAT} Collaboration, M.~Ackermann {\em et~al.}, ``{Searching
  for Dark Matter Annihilation from Milky Way Dwarf Spheroidal Galaxies with
  Six Years of Fermi Large Area Telescope Data},''
  \href{http://dx.doi.org/10.1103/PhysRevLett.115.231301}{{\em Phys. Rev.
  Lett.} {\bfseries 115} no.~23, (2015) 231301},
\href{http://arxiv.org/abs/1503.02641}{{\ttfamily arXiv:1503.02641
  [astro-ph.HE]}}.

\bibitem{Charles:2016pgz}
{\bfseries Fermi-LAT} Collaboration, E.~Charles {\em et~al.}, ``{Sensitivity
  Projections for Dark Matter Searches with the Fermi Large Area Telescope},''
  \href{http://dx.doi.org/10.1016/j.physrep.2016.05.001}{{\em Phys. Rept.}
  {\bfseries 636} (2016) 1--46},
\href{http://arxiv.org/abs/1605.02016}{{\ttfamily arXiv:1605.02016
  [astro-ph.HE]}}.

\bibitem{Aleksic:2011jx}
{\bfseries MAGIC} Collaboration, J.~Aleksic {\em et~al.}, ``{Searches for Dark
  Matter annihilation signatures in the Segue 1 satellite galaxy with the
  MAGIC-I telescope},''
  \href{http://dx.doi.org/10.1088/1475-7516/2011/06/035}{{\em JCAP} {\bfseries
  1106} (2011) 035},
\href{http://arxiv.org/abs/1103.0477}{{\ttfamily arXiv:1103.0477
  [astro-ph.HE]}}.

\bibitem{Aleksic:2013xea}
J.~Aleksi{\'c} {\em et~al.}, ``{Optimized dark matter searches in deep
  observations of Segue 1 with MAGIC},''
  \href{http://dx.doi.org/10.1088/1475-7516/2014/02/008}{{\em JCAP} {\bfseries
  1402} (2014) 008},
\href{http://arxiv.org/abs/1312.1535}{{\ttfamily arXiv:1312.1535 [hep-ph]}}.

\bibitem{Doro:2017dqn}
{\bfseries MAGIC} Collaboration, M.~Doro, ``{A review of the past and present
  MAGIC dark matter search program and a glimpse at the future},'' in {\em
  {25th European Cosmic Ray Symposium (ECRS 2016) Turin, Italy, September
  04-09, 2016}}.
\newblock 2017.
\newblock \href{http://arxiv.org/abs/1701.05702}{{\ttfamily arXiv:1701.05702
  [astro-ph.HE]}}.
\newblock
\url{https://inspirehep.net/record/1510040/files/arXiv:1701.05702.pdf}.
\newblock

\bibitem{Ahnen:2017pqx}
{\bfseries MAGIC} Collaboration, M.~L. Ahnen {\em et~al.}, ``{Indirect dark
  matter searches in the dwarf satellite galaxy Ursa Major II with the MAGIC
  Telescopes},'' \href{http://dx.doi.org/10.1088/1475-7516/2018/03/009}{{\em
  JCAP} {\bfseries 1803} no.~03, (2018) 009},
\href{http://arxiv.org/abs/1712.03095}{{\ttfamily arXiv:1712.03095
  [astro-ph.HE]}}.

\bibitem{Aharonian:2007km}
{\bfseries H.E.S.S.} Collaboration, F.~Aharonian, ``{Observations of the
  Sagittarius Dwarf galaxy by the H.E.S.S. experiment and search for a Dark
  Matter signal},''
  \href{http://dx.doi.org/10.1016/j.astropartphys.2007.11.007,
  10.1016/j.astropartphys.2010.01.007}{{\em Astropart. Phys.} {\bfseries 29}
  (2008) 55--62}, \href{http://arxiv.org/abs/0711.2369}{{\ttfamily
  arXiv:0711.2369 [astro-ph]}}.
[Erratum: Astropart. Phys.33,274(2010)].

\bibitem{Aharonian:2008dm}
{\bfseries H.E.S.S.} Collaboration, F.~Aharonian, ``{A search for a dark matter
  annihilation signal towards the Canis Major overdensity with H.E.S.S},''
  \href{http://dx.doi.org/10.1088/0004-637X/691/1/175}{{\em Astrophys. J.}
  {\bfseries 691} (2009) 175--181},
\href{http://arxiv.org/abs/0809.3894}{{\ttfamily arXiv:0809.3894 [astro-ph]}}.

\bibitem{Abramowski:2010aa}
{\bfseries H.E.S.S.} Collaboration, A.~Abramowski {\em et~al.}, ``{H.E.S.S.
  constraints on Dark Matter annihilations towards the Sculptor and Carina
  Dwarf Galaxies},''
  \href{http://dx.doi.org/10.1016/j.astropartphys.2010.12.006}{{\em Astropart.
  Phys.} {\bfseries 34} (2011) 608--616},
\href{http://arxiv.org/abs/1012.5602}{{\ttfamily arXiv:1012.5602
  [astro-ph.HE]}}.

\bibitem{Abramowski:2014tra}
{\bfseries H.E.S.S.} Collaboration, A.~Abramowski {\em et~al.}, ``{Search for
  dark matter annihilation signatures in H.E.S.S. observations of Dwarf
  Spheroidal Galaxies},''
  \href{http://dx.doi.org/10.1103/PhysRevD.90.112012}{{\em Phys. Rev.}
  {\bfseries D90} (2014) 112012},
\href{http://arxiv.org/abs/1410.2589}{{\ttfamily arXiv:1410.2589
  [astro-ph.HE]}}.

\bibitem{Acciari:2010ab}
{\bfseries VERITAS} Collaboration, V.~A. Acciari {\em et~al.}, ``{VERITAS
  Search for VHE Gamma-ray Emission from Dwarf Spheroidal Galaxies},''
  \href{http://dx.doi.org/10.1088/0004-637X/720/2/1174}{{\em Astrophys. J.}
  {\bfseries 720} (2010) 1174--1180},
\href{http://arxiv.org/abs/1006.5955}{{\ttfamily arXiv:1006.5955
  [astro-ph.CO]}}.

\bibitem{Aliu:2012ga}
{\bfseries VERITAS} Collaboration, E.~Aliu {\em et~al.}, ``{VERITAS Deep
  Observations of the Dwarf Spheroidal Galaxy Segue 1},''
  \href{http://dx.doi.org/10.1103/PhysRevD.85.062001,
  10.1103/PhysRevD.91.129903}{{\em Phys. Rev.} {\bfseries D85} (2012) 062001},
  \href{http://arxiv.org/abs/1202.2144}{{\ttfamily arXiv:1202.2144
  [astro-ph.HE]}}.
[Erratum: Phys. Rev.D91,no.12,129903(2015)].

\bibitem{Archambault:2017wyh}
{\bfseries VERITAS} Collaboration, S.~Archambault {\em et~al.}, ``{Dark Matter
  Constraints from a Joint Analysis of Dwarf Spheroidal Galaxy Observations
  with VERITAS},'' \href{http://dx.doi.org/10.1103/PhysRevD.95.082001}{{\em
  Phys. Rev.} {\bfseries D95} no.~8, (2017) 082001},
\href{http://arxiv.org/abs/1703.04937}{{\ttfamily arXiv:1703.04937
  [astro-ph.HE]}}.

\bibitem{Albert:2017vtb}
{\bfseries HAWC} Collaboration, A.~Albert {\em et~al.}, ``{Dark Matter Limits
  From Dwarf Spheroidal Galaxies with The HAWC Gamma-Ray Observatory},''
  \href{http://dx.doi.org/10.3847/1538-4357/aaa6d8}{{\em Astrophys. J.}
  {\bfseries 853} no.~2, (2018) 154},
\href{http://arxiv.org/abs/1706.01277}{{\ttfamily arXiv:1706.01277
  [astro-ph.HE]}}.

\bibitem{Essig:2010em}
R.~Essig, N.~Sehgal, L.~E. Strigari, M.~Geha, and J.~D. Simon, ``{Indirect Dark
  Matter Detection Limits from the Ultra-Faint Milky Way Satellite Segue 1},''
  \href{http://dx.doi.org/10.1103/PhysRevD.82.123503}{{\em Phys. Rev.}
  {\bfseries D82} (2010) 123503},
\href{http://arxiv.org/abs/1007.4199}{{\ttfamily arXiv:1007.4199
  [astro-ph.CO]}}.

\bibitem{Ahnen:2016qkx}
{\bfseries Fermi-LAT, MAGIC} Collaboration, M.~L. Ahnen {\em et~al.}, ``{Limits
  to dark matter annihilation cross-section from a combined analysis of MAGIC
  and Fermi-LAT observations of dwarf satellite galaxies},''
  \href{http://dx.doi.org/10.1088/1475-7516/2016/02/039}{{\em JCAP} {\bfseries
  1602} no.~02, (2016) 039},
\href{http://arxiv.org/abs/1601.06590}{{\ttfamily arXiv:1601.06590
  [astro-ph.HE]}}.

\bibitem{Ambrogi:2018skq}
L.~Ambrogi, S.~Celli, and F.~Aharonian, ``{On the potential of Cherenkov
  Telescope Arrays and KM3 Neutrino Telescopes for the detection of extended
  sources},'' \href{http://dx.doi.org/10.1016/j.astropartphys.2018.03.001}{{\em
  Astropart. Phys.} {\bfseries 100} (2018) 69--79},
\href{http://arxiv.org/abs/1803.03565}{{\ttfamily arXiv:1803.03565
  [astro-ph.HE]}}.

\bibitem{Bonnivard:2015vua}
V.~Bonnivard, D.~Maurin, and M.~G. Walker, ``{Contamination of
  stellar-kinematic samples and uncertainty about dark matter annihilation
  profiles in ultrafaint dwarf galaxies: the example of Segue I},''
  \href{http://dx.doi.org/10.1093/mnras/stw1691}{{\em Mon. Not. Roy. Astron.
  Soc.} {\bfseries 462} no.~1, (2016) 223--234},
\href{http://arxiv.org/abs/1506.08209}{{\ttfamily arXiv:1506.08209
  [astro-ph.GA]}}.

\bibitem{Mashchenko:2005bj}
S.~Mashchenko, A.~Sills, and H.~M.~P. Couchman, ``{Constraining global
  properties of the draco dwarf spheroidal galaxy},''
  \href{http://dx.doi.org/10.1086/499940}{{\em Astrophys. J.} {\bfseries 640}
  (2006) 252--269},
\href{http://arxiv.org/abs/astro-ph/0511567}{{\ttfamily arXiv:astro-ph/0511567
  [astro-ph]}}.

\bibitem{SanchezConde:2007te}
M.~A. Sanchez-Conde, F.~Prada, E.~L. Lokas, M.~E. Gomez, R.~Wojtak, and
  M.~Moles, ``{Dark Matter annihilation in Draco: New considerations of the
  expected gamma flux},''
  \href{http://dx.doi.org/10.1103/PhysRevD.76.123509}{{\em Phys. Rev.}
  {\bfseries D76} (2007) 123509},
\href{http://arxiv.org/abs/astro-ph/0701426}{{\ttfamily arXiv:astro-ph/0701426
  [astro-ph]}}.

\bibitem{Geringer-Sameth:2014yza}
A.~Geringer-Sameth, S.~M. Koushiappas, and M.~Walker, ``{Dwarf galaxy
  annihilation and decay emission profiles for dark matter experiments},''
  \href{http://dx.doi.org/10.1088/0004-637X/801/2/74}{{\em Astrophys. J.}
  {\bfseries 801} no.~2, (2015) 74},
\href{http://arxiv.org/abs/1408.0002}{{\ttfamily arXiv:1408.0002
  [astro-ph.CO]}}.

\bibitem{Lokas:2001mf}
E.~L. Lokas, ``{Dark matter distribution in dwarf spheroidal galaxies},''
  \href{http://dx.doi.org/10.1046/j.1365-8711.2002.05457.x}{{\em Mon. Not. Roy.
  Astron. Soc.} {\bfseries 333} (2002) 697},
\href{http://arxiv.org/abs/astro-ph/0112023}{{\ttfamily arXiv:astro-ph/0112023
  [astro-ph]}}.

\bibitem{Lokas:2004sw}
E.~L. Lokas, G.~A. Mamon, and F.~Prada, ``{Dark matter distribution in the
  Draco dwarf from velocity moments},''
  \href{http://dx.doi.org/10.1111/j.1365-2966.2005.09497.x}{{\em Mon. Not. Roy.
  Astron. Soc.} {\bfseries 363} (2005) 918},
\href{http://arxiv.org/abs/astro-ph/0411694}{{\ttfamily arXiv:astro-ph/0411694
  [astro-ph]}}.

\bibitem{Knodlseder:2016nnv}
J.~Kn{\"o}dlseder {\em et~al.}, ``{GammaLib and ctools: A software framework
  for the analysis of astronomical gamma-ray data},''
  \href{http://dx.doi.org/10.1051/0004-6361/201628822}{{\em Astron. Astrophys.}
  {\bfseries 593} (2016) A1},
\href{http://arxiv.org/abs/1606.00393}{{\ttfamily arXiv:1606.00393
  [astro-ph.IM]}}.

\bibitem{Hernquist:1990be}
L.~Hernquist, ``{An Analytical Model for Spherical Galaxies and Bulges},''
\href{http://dx.doi.org/10.1086/168845}{{\em Astrophys. J.} {\bfseries 356}
  (1990) 359}.

\bibitem{Zhao:1995cp}
H.~Zhao, ``{Analytical models for galactic nuclei},''
  \href{http://dx.doi.org/10.1093/mnras/278.2.488}{{\em Mon. Not. Roy. Astron.
  Soc.} {\bfseries 278} (1996) 488--496},
\href{http://arxiv.org/abs/astro-ph/9509122}{{\ttfamily arXiv:astro-ph/9509122
  [astro-ph]}}.

\bibitem{Navarro:1996gj}
J.~F. Navarro, C.~S. Frenk, and S.~D.~M. White, ``{A Universal density profile
  from hierarchical clustering},'' \href{http://dx.doi.org/10.1086/304888}{{\em
  Astrophys. J.} {\bfseries 490} (1997) 493--508},
\href{http://arxiv.org/abs/astro-ph/9611107}{{\ttfamily arXiv:astro-ph/9611107
  [astro-ph]}}.

\bibitem{Burkert:1995yz}
A.~Burkert, ``{The Structure of dark matter halos in dwarf galaxies},''
  \href{http://dx.doi.org/10.1086/309560}{{\em IAU Symp.} {\bfseries 171}
  (1996) 175}, \href{http://arxiv.org/abs/astro-ph/9504041}{{\ttfamily
  arXiv:astro-ph/9504041 [astro-ph]}}.
[Astrophys. J.447,L25(1995)].

\bibitem{Bonnivard:2014kza}
V.~Bonnivard, C.~Combet, D.~Maurin, and M.~G. Walker, ``{Spherical Jeans
  analysis for dark matter indirect detection in dwarf spheroidal galaxies -
  Impact of physical parameters and triaxiality},''
  \href{http://dx.doi.org/10.1093/mnras/stu2296}{{\em Mon. Not. Roy. Astron.
  Soc.} {\bfseries 446} (2015) 3002--3021},
\href{http://arxiv.org/abs/1407.7822}{{\ttfamily arXiv:1407.7822
  [astro-ph.HE]}}.

\bibitem{Bonnivard:2015pia}
V.~Bonnivard, M.~H{\"u}tten, E.~Nezri, A.~Charbonnier, C.~Combet, and
  D.~Maurin, ``{CLUMPY : Jeans analysis, $\gamma$-ray and $\nu$ fluxes from
  dark matter (sub-)structures},''
  \href{http://dx.doi.org/10.1016/j.cpc.2015.11.012}{{\em Comput. Phys.
  Commun.} {\bfseries 200} (2016) 336--349},
\href{http://arxiv.org/abs/1506.07628}{{\ttfamily arXiv:1506.07628
  [astro-ph.CO]}}.

\bibitem{Hutten:2016jko}
M.~H{\"u}tten, C.~Combet, G.~Maier, and D.~Maurin, ``{Dark matter substructure
  modelling and sensitivity of the Cherenkov Telescope Array to Galactic dark
  halos},'' \href{http://dx.doi.org/10.1088/1475-7516/2016/09/047}{{\em JCAP}
  {\bfseries 1609} no.~09, (2016) 047},
\href{http://arxiv.org/abs/1606.04898}{{\ttfamily arXiv:1606.04898
  [astro-ph.HE]}}.

\bibitem{Sjostrand:2014zea}
T.~Sj{\"o}strand, S.~Ask, J.~R. Christiansen, R.~Corke, N.~Desai, P.~Ilten,
  S.~Mrenna, S.~Prestel, C.~O. Rasmussen, and P.~Z. Skands, ``{An Introduction
  to PYTHIA 8.2},'' \href{http://dx.doi.org/10.1016/j.cpc.2015.01.024}{{\em
  Comput. Phys. Commun.} {\bfseries 191} (2015) 159--177},
\href{http://arxiv.org/abs/1410.3012}{{\ttfamily arXiv:1410.3012 [hep-ph]}}.

\bibitem{Sjostrand:2006za}
T.~Sjostrand, S.~Mrenna, and P.~Z. Skands, ``{PYTHIA 6.4 Physics and Manual},''
  \href{http://dx.doi.org/10.1088/1126-6708/2006/05/026}{{\em JHEP} {\bfseries
  05} (2006) 026},
\href{http://arxiv.org/abs/hep-ph/0603175}{{\ttfamily arXiv:hep-ph/0603175
  [hep-ph]}}.

\bibitem{Sjostrand:2007gs}
T.~Sjostrand, S.~Mrenna, and P.~Z. Skands, ``{A Brief Introduction to PYTHIA
  8.1},'' \href{http://dx.doi.org/10.1016/j.cpc.2008.01.036}{{\em Comput. Phys.
  Commun.} {\bfseries 178} (2008) 852--867},
\href{http://arxiv.org/abs/0710.3820}{{\ttfamily arXiv:0710.3820 [hep-ph]}}.

\bibitem{Belikov:2009cx}
A.~V. Belikov and D.~Hooper, ``{The Contribution Of Inverse Compton Scattering
  To The Diffuse Extragalactic Gamma-Ray Background From Annihilating Dark
  Matter},'' \href{http://dx.doi.org/10.1103/PhysRevD.81.043505}{{\em Phys.
  Rev.} {\bfseries D81} (2010) 043505},
\href{http://arxiv.org/abs/0906.2251}{{\ttfamily arXiv:0906.2251
  [astro-ph.CO]}}.

\bibitem{Cirelli:2009vg}
M.~Cirelli and P.~Panci, ``{Inverse Compton constraints on the Dark Matter e+e-
  excesses},'' \href{http://dx.doi.org/10.1016/j.nuclphysb.2009.06.034}{{\em
  Nucl. Phys.} {\bfseries B821} (2009) 399--416},
\href{http://arxiv.org/abs/0904.3830}{{\ttfamily arXiv:0904.3830
  [astro-ph.CO]}}.

\bibitem{Profumo:2009uf}
S.~Profumo and T.~E. Jeltema, ``{Extragalactic Inverse Compton Light from Dark
  Matter Annihilation and the Pamela Positron Excess},''
  \href{http://dx.doi.org/10.1088/1475-7516/2009/07/020}{{\em JCAP} {\bfseries
  0907} (2009) 020},
\href{http://arxiv.org/abs/0906.0001}{{\ttfamily arXiv:0906.0001
  [astro-ph.CO]}}.

\bibitem{Blanco:2017sbc}
C.~Blanco, J.~P. Harding, and D.~Hooper, ``{Novel Gamma-Ray Signatures of
  PeV-Scale Dark Matter},''
  \href{http://dx.doi.org/10.1088/1475-7516/2018/04/060}{{\em JCAP} {\bfseries
  1804} no.~04, (2018) 060},
\href{http://arxiv.org/abs/1712.02805}{{\ttfamily arXiv:1712.02805 [hep-ph]}}.

\bibitem{Bartels:2017dpb}
R.~Bartels, D.~Gaggero, and C.~Weniger, ``{Prospects for indirect dark matter
  searches with MeV photons},''
  \href{http://dx.doi.org/10.1088/1475-7516/2017/05/001}{{\em JCAP} {\bfseries
  1705} no.~05, (2017) 001},
\href{http://arxiv.org/abs/1703.02546}{{\ttfamily arXiv:1703.02546
  [astro-ph.HE]}}.

\bibitem{Cirelli:2010xx}
M.~Cirelli, G.~Corcella, A.~Hektor, G.~Hutsi, M.~Kadastik, P.~Panci, M.~Raidal,
  F.~Sala, and A.~Strumia, ``{PPPC 4 DM ID: A Poor Particle Physicist Cookbook
  for Dark Matter Indirect Detection},''
  \href{http://dx.doi.org/10.1088/1475-7516/2012/10/E01,
  10.1088/1475-7516/2011/03/051}{{\em JCAP} {\bfseries 1103} (2011) 051},
  \href{http://arxiv.org/abs/1012.4515}{{\ttfamily arXiv:1012.4515 [hep-ph]}}.
[Erratum: JCAP1210,E01(2012)].

\bibitem{Cumani:2017aca}
{\bfseries CTA Consortium} Collaboration, P.~Cumani, T.~Hassan, L.~Arrabito,
  K.~Bernl{\"o}hr, J.~Bregeon, G.~Maier, and A.~Moralejo, ``{Baseline telescope
  layouts of the Cherenkov Telescope Array},''
  \href{http://dx.doi.org/10.22323/1.301.0811}{{\em PoS} {\bfseries ICRC2017}
  (2018) 811},
\href{http://arxiv.org/abs/1709.00206}{{\ttfamily arXiv:1709.00206
  [astro-ph.IM]}}.

\bibitem{Abdalla:2018mve}
{\bfseries HESS} Collaboration, H.~Abdalla {\em et~al.}, ``{Searches for
  gamma-ray lines and 'pure WIMP' spectra from Dark Matter annihilations in
  dwarf galaxies with H.E.S.S},''
  \href{http://dx.doi.org/10.1088/1475-7516/2018/11/037}{{\em JCAP} {\bfseries
  1811} no.~11, (2018) 037},
\href{http://arxiv.org/abs/1810.00995}{{\ttfamily arXiv:1810.00995
  [astro-ph.HE]}}.

\bibitem{Wood:2013taa}
M.~Wood, J.~Buckley, S.~Digel, S.~Funk, D.~Nieto, and M.~A. Sanchez-Conde,
  ``{Prospects for Indirect Detection of Dark Matter with CTA},'' in {\em
  {Proceedings, 2013 Community Summer Study on the Future of U.S. Particle
  Physics: Snowmass on the Mississippi (CSS2013): Minneapolis, MN, USA, July
  29-August 6, 2013}}.
\newblock 2013.
\newblock \href{http://arxiv.org/abs/1305.0302}{{\ttfamily arXiv:1305.0302
  [astro-ph.HE]}}.
\newblock
\url{http://www.slac.stanford.edu/econf/C1307292/docs/submittedArxivFiles/1305.0302.pdf}.
\newblock

\bibitem{Silverwood:2014yza}
H.~Silverwood, C.~Weniger, P.~Scott, and G.~Bertone, ``{A realistic assessment
  of the CTA sensitivity to dark matter annihilation},''
  \href{http://dx.doi.org/10.1088/1475-7516/2015/03/055}{{\em JCAP} {\bfseries
  1503} no.~03, (2015) 055},
\href{http://arxiv.org/abs/1408.4131}{{\ttfamily arXiv:1408.4131
  [astro-ph.HE]}}.

\bibitem{Roszkowski:2014iqa}
L.~Roszkowski, E.~M. Sessolo, and A.~J. Williams, ``{Prospects for dark matter
  searches in the pMSSM},''
  \href{http://dx.doi.org/10.1007/JHEP02(2015)014}{{\em JHEP} {\bfseries 02}
  (2015) 014},
\href{http://arxiv.org/abs/1411.5214}{{\ttfamily arXiv:1411.5214 [hep-ph]}}.

\bibitem{Sanchez-Conde:2013yxa}
M.~A. Sanchez-Conde and F.~Prada, ``{The flattening of the concentration-mass
  relation towards low halo masses and its implications for the annihilation
  signal boost},'' \href{http://dx.doi.org/10.1093/mnras/stu1014}{{\em Mon.
  Not. Roy. Astron. Soc.} {\bfseries 442} no.~3, (2014) 2271--2277},
\href{http://arxiv.org/abs/1312.1729}{{\ttfamily arXiv:1312.1729
  [astro-ph.CO]}}.

\bibitem{Moline:2016pbm}
A.~Moline, M.~A. Sanchez-Conde, S.~Palomares-Ruiz, and F.~Prada,
  ``{Characterization of subhalo structural properties and implications for
  dark matter annihilation signals},''
  \href{http://dx.doi.org/10.1093/mnras/stx026}{{\em Mon. Not. Roy. Astron.
  Soc.} {\bfseries 466} no.~4, (2017) 4974--4990},
\href{http://arxiv.org/abs/1603.04057}{{\ttfamily arXiv:1603.04057
  [astro-ph.CO]}}.

\bibitem{Stref:2016uzb}
M.~Stref and J.~Lavalle, ``{Modeling dark matter subhalos in a constrained
  galaxy: Global mass and boosted annihilation profiles},''
  \href{http://dx.doi.org/10.1103/PhysRevD.95.063003}{{\em Phys. Rev.}
  {\bfseries D95} no.~6, (2017) 063003},
\href{http://arxiv.org/abs/1610.02233}{{\ttfamily arXiv:1610.02233
  [astro-ph.CO]}}.

\bibitem{Hiroshima:2018kfv}
N.~Hiroshima, S.~Ando, and T.~Ishiyama, ``{Modeling evolution of dark matter
  substructure and annihilation boost},''
  \href{http://dx.doi.org/10.1103/PhysRevD.97.123002}{{\em Phys. Rev.}
  {\bfseries D97} no.~12, (2018) 123002},
\href{http://arxiv.org/abs/1803.07691}{{\ttfamily arXiv:1803.07691
  [astro-ph.CO]}}.

\bibitem{Charbonnier:2012gf}
A.~Charbonnier, C.~Combet, and D.~Maurin, ``{CLUMPY: a code for gamma-ray
  signals from dark matter structures},''
  \href{http://dx.doi.org/10.1016/j.cpc.2011.10.017}{{\em Comput. Phys.
  Commun.} {\bfseries 183} (2012) 656--668},
\href{http://arxiv.org/abs/1201.4728}{{\ttfamily arXiv:1201.4728
  [astro-ph.HE]}}.

\bibitem{Hutten:2018aix}
M.~Hutten, C.~Combet, and D.~Maurin, ``{CLUMPY v3: $\gamma$-ray and $\nu$
  signals from dark matter at all scales},''
  \href{http://dx.doi.org/10.1016/j.cpc.2018.10.001}{{\em Comput. Phys.
  Commun.} {\bfseries 235} (2019) 336--345},
\href{http://arxiv.org/abs/1806.08639}{{\ttfamily arXiv:1806.08639
  [astro-ph.CO]}}.

\bibitem{Evans:2016xwx}
N.~W. Evans, J.~L. Sanders, and A.~Geringer-Sameth, ``{Simple J-Factors and
  D-Factors for Indirect Dark Matter Detection},''
  \href{http://dx.doi.org/10.1103/PhysRevD.93.103512}{{\em Phys. Rev.}
  {\bfseries D93} no.~10, (2016) 103512},
\href{http://arxiv.org/abs/1604.05599}{{\ttfamily arXiv:1604.05599
  [astro-ph.GA]}}.

\bibitem{Hassan:2017paq}
T.~Hassan {\em et~al.}, ``{Monte Carlo Performance Studies for the Site
  Selection of the Cherenkov Telescope Array},''
  \href{http://dx.doi.org/10.1016/j.astropartphys.2017.05.001}{{\em Astropart.
  Phys.} {\bfseries 93} (2017) 76--85},
\href{http://arxiv.org/abs/1705.01790}{{\ttfamily arXiv:1705.01790
  [astro-ph.IM]}}.

\bibitem{Calore:2018sdx}
F.~Calore, P.~D. Serpico, and B.~Zaldivar, ``{Dark matter constraints from
  dwarf galaxies: a data-driven analysis},''
  \href{http://dx.doi.org/10.1088/1475-7516/2018/10/029}{{\em JCAP} {\bfseries
  1810} no.~10, (2018) 029},
\href{http://arxiv.org/abs/1803.05508}{{\ttfamily arXiv:1803.05508
  [astro-ph.HE]}}.

\end{thebibliography}\endgroup
\end{document}